\documentclass[11pt]{article}

\usepackage[english]{babel}

\newcommand{\proglang}{\texttt}
\newcommand{\pkg}{\textbf}
\newcommand{\code}{\texttt}

\usepackage{natbib}
\usepackage{xcolor}
\usepackage{soul}
\usepackage{amsfonts}
\usepackage{amsmath,amssymb}
\usepackage{graphicx}
\usepackage{appendix}
\usepackage{amsthm}
\usepackage{hyperref}

\newcommand{\R}{\mathbb{R}}
\newcommand{\I}{\mathbb{I}}

\newcommand{\bZ}{\textbf{Z}}
\newcommand{\bX}{\textbf{X}}
\newcommand{\bM}{\textbf{M}}
\newcommand{\bN}{\textbf{N}}
\newcommand{\bLambda}{\boldsymbol{\Lambda}}
\newcommand{\bU}{\textbf{U}}
\newcommand{\bV}{\textbf{V}}
\newcommand{\bF}{\textbf{F}}
\newcommand{\bA}{\textbf{A}}
\newcommand{\bG}{\textbf{G}}

\newcommand{\bT}{\textbf{T}}
\newcommand{\bP}{\textbf{P}}

\newcommand{\tbU}{{\tilde \bU}}
\newcommand{\tbV}{{\tilde \bV}}
\newcommand{\tbLambda}{{\tilde \bLambda}}
\newcommand{\tbZ}{{\tilde \bZ}}
\newcommand{\tbA}{{\tilde \bA}}
\newcommand{\tbF}{{\tilde \bF}}

\newcommand{\bz}{\textbf{z}}
\newcommand{\bx}{\textbf{x}}
\newcommand{\ba}{\textbf{a}}

\newcommand{\bv}{\textbf{v}}
\newcommand{\bu}{\textbf{u}}
\newcommand{\bff}{\textbf{f}}

\newcommand{\tr}{\mathrm{tr}}

\newcommand{\ta}{{\tilde a}}

\newcommand{\tba}{{\tilde \ba}}

\newcommand{\g}{(g)}
\newcommand{\MFA}{{\mbox{\tiny mfa}}}
\newcommand{\pa}{{\mbox{\tiny part}}}

\newtheorem{remark}{Remark}

\title{Multivariate Analysis of Mixed Data. The R Package PCAmixdata}

\author{Marie Chavent$^{1}$, Vanessa Kuentz$^{2}$, Amaury Labenne$^{2}$,  J\'er\^ome Saracco$^{1}$}

\date{{\small \today }}

\begin{document}

\maketitle 

\begin{center}
\small $^1$ Univ. Bordeaux, CNRS, INRIA, Bordeaux INP, IMB, UMR 5251, F-33400 Talence, France\\
  \smallskip
 $^2$ INRAE, ETTIS, F-33612 Cestas, France\\
\end{center}

\begin{abstract}
Mixed data  arise when observations are described by a mixture of numerical and categorical variables. The \proglang{R} package \pkg{PCAmixdata} extends to this type of data standard multivariate analysis methods which allow description, exploration and visualization of the data. The key techniques/methods included in the package are principal component analysis for mixed data (\code{PCAmix}), varimax-like orthogonal rotation for \code{PCAmix},  and multiple factor analysis for mixed multi-table data. This paper proposes a unified mathematical presentation of the different methods with common notations, as well as providing a summarised presentation of the three algorithms, with details to help the user understand graphical and numerical outputs of the corresponding \proglang{R} functions. 
This then allows the user to easily provide relevant interpretations of the results obtained. 
The three main methods are illustrated  on a real dataset composed of four data tables characterizing living conditions in different municipalities in the Gironde region of southwest France.

\paragraph{keywords:} mixture of numerical and categorical data, PCA, multiple correspondence analysis, multiple factor analysis,  varimax rotation, \proglang{R}.
\end{abstract}

\section{Introduction}

Multivariate data analysis refers  to descriptive statistical methods used to analyze data arising from more than one variable. 
The common goal of these methods is to provide description, exploration and visualization of the data, via data reduction and graphical display.
These variables can be either numerical or categorical. For example, principal component analysis (PCA) and varimax rotation handle numerical variables, whereas multiple correspondence analysis (MCA) handles categorical variables. Multiple factor analysis (MFA) works with multi-table data, where the type of the variables can vary from one data table to the other but the variables should be of the same type within a given data table \citep{escofier1994multiple, becue2008, abdi2013multiple}. 

While PCA, varimax and MFA for mixed data have been already described elsewhere, our paper provides a unified mathematical presentation of all the different methods with common notations, which greatly enhances the user experience. A synthetic presentation of the corresponding algorithms is given, with details to help the user understand all the graphical and numerical outputs of the \proglang{R}  package \pkg{PCAmixdata} \citep{pcamixdatar}. The way the methods are presented is clearly in the French tradition of data analysis (``analyse des données'' in French), following the words of Jean-Paul Benzecri ``all in all, doing a data analysis, in good mathematics, is simply searching for eigenvectors; all the science (or the art) of it is in finding the right matrix to diagonalize''. More precisely, the underlying theory involves reducing data dimensionality, via generalized singular value decomposition, to provide a subspace that best represents the data in the sense of maximizing the variability of the projected points. Because of this, great importance is attached to relevant graphical representations of both rows and columns of the data matrices.

Several existing \proglang{R}  \citep{team2017r} packages use standard multivariate analysis methods. These include \pkg{ade4} \citep{dray2007ade4,rade4}, \pkg{FactoMineR} \citep{le2008factominer,rFactoMineR}, \pkg{ExPosition} \citep{pExPosition} or \pkg{Gifi} \citep{gifi, homals}. Most of them propose a function to perform PCA with a mixture of numerical and categorical data. For instance, the function \code{dudi.mix} in the package \pkg{ade4} implements the method  developed by \cite{hill1976principal} and the function \code{FAMD} of the package \pkg{FactoMineR} implements that developed by \cite{pages2004analyse} in the spirit of the French school. Note that these methods are also equivalent to PCAMIX, a method proposed independently by \cite{deleeuw80}  and extended by \cite{kiers1991simple}, in the spirit of the Dutch school. The \proglang{R}  package \pkg{PCAmixdata} presented in this paper is dedicated to mixed data and provides three main functions: 
\begin{itemize}
\item \code{PCAmix} (PCA of a mixture of numerical and categorical variables),  
\item \code{PCArot} (rotation after \code{PCAmix}), 
\item and  \code{MFAmix} (multiple factor analysis of mixed multi-table data).
\end{itemize}

The  \code{PCAmix} function gives similar results to \code{dudi.mix}  and  \code{FAMD}. The procedure  \code{PCArot} \citep{chavent2012orthogonal}  is not implemented elsewhere and allows in particular to make rotation in MCA for categorical data. The function  \code{MFAmix} allows mixed single data table (groups with both numerical and categorical variables) and differs  from the function \code{MFA} of the package \pkg{FactoMineR}  where only groups of numerical and groups of categorical variables are allowed.

In addition, the package  \pkg{PCAmixdata} naturally offers functions to plot graphical outputs. One particularly useful feature of the package is the possibility to predict scores for new observations of the principal components of \code{PCAmix}, \code{PCArot} and \code{MFAmix}, and to project supplementary variables or levels (resp. supplementary groups of variables) on the maps of \code{PCAmix} (resp.  \code{MFAmix}). These functions are implemented in the \proglang{R} package as S3 methods with generic names \code{plot}, \code{predict} and \code{suppvar} associated with the objects of class \code{PCAmix}, \code{PCArot} and \code{MFAmix}. 

A real dataset called \code{gironde} is available in the package to illustrate the functions and the outputs with simple examples. This will allow the user to easily understand the numerical and graphical outputs, and thus  draw fine and relevant interpretations from the results obtained. This dataset is made up of four data tables, each characterizing living conditions in $542$ municipalities in the Gironde region in southwest France, see Appendix~\ref{datades} for details. This dataset was taken from the 2009 census database\footnote{http://www.insee.fr/fr/bases-de-donnees/} of the French national institute of statistics and economic studies and from a topographic database\footnote{http://professionnels.ign.fr/bdtopo} of the French national institute of geographic and forestry information. The first data table describes  the $542$ municipalities  with 9 numerical variables relating to employment conditions. The second data table describes those municipalities with  5 variables (2 categorical and 3 numerical) relating to housing conditions, the third one with 9 categorical variables relating to services (restaurants, doctors, post offices,...) and the last one with 4 numerical variables relating to environmental conditions. A complete description of the 27 variables, divided into 4 groups (Employment, Housing, Services, Environment) is given in Appendix \ref{datades}.

The rest of the paper is organized as follows. Section \ref{section_gsvd} details the link between standard PCA and MCA via Generalized Singular Value Decomposition (GSVD). It demonstrates how MCA can be obtained from a single PCA with metrics, the cornerstone for merging standard PCA and MCA in \code{PCAmix}. Sections \ref{pcamix_section}, \ref{pcarot_section} and \ref{mfamix_section}  present respectively the   \code{PCAmix},   \code{PCArot} and  \code{MFAmix} methods with details for the interpretation of the associated graphical and numerical outputs. 
Some technical proofs have been provided in Appendices~\ref{Appendix:pcarot} and \ref{appendix:proof}.
In each of these sections, the corresponding method is illustrated with the \code{gironde} dataset  and the associated \proglang{R} code is given. 
Finally, concluding remarks are provided in Section~\ref{conclu}.

\section{PCA with metrics} \label{section_gsvd}

PCA with metrics is a generalization of the standard PCA method where metrics are used to introduce weights into rows (observations) and columns (variables) within a data matrix. Standard PCA for numerical data and standard MCA for categorical data can be presented within this general framework, so that the unique  PCAmix procedure for a mixture of numerical and categorical data can be easily  defined.

\subsection{The general framework} \label{sec:pcamet}
Let $\textbf{Z}$ be a real matrix of dimension $n \times p$.  Let  $\bN$ (resp. $\bM$) be the diagonal matrix of the weights of the  $n$ rows (resp. the weights of the $p$ columns).  

\paragraph{Generalized Singular Value Decomposition.} 

The GSVD of $\bZ$ with metrics $\bN$ on $\mathbb{R}^n$ and $\bM$  on $\mathbb{R}^p$ gives the following decomposition:
\begin{equation} \label{gsvd}
\bZ=\bU \bLambda  \bV^\top,
\end{equation}

where
\begin{itemize}
\item[-] $\bLambda=\mbox{diag}(\sqrt{\lambda_1},\ldots,\sqrt{\lambda_r})$ is the $r\times r$ diagonal matrix of the singular values of $\bZ\bM\bZ^\top\bN$ and $\bZ^\top\bN\bZ\bM$, and $r$ denotes the rank of $\bZ$;
\item[-] $\bU$ is the $n \times r$  matrix of the first $r$ eigenvectors of $\bZ \bM \bZ^\top \bN$ such that $\bU^\top \bN \bU=\mathbb{I}_r$, with $\mathbb{I}_r$ the identity matrix of size $n \times r$;
\item[-] $\bV$ is the $p \times r$  matrix of the first $r$ eigenvectors of $\bZ^\top\bN\bZ\bM$ such that $\bV^\top \bM \bV=\mathbb{I}_r$.
\end{itemize}

\begin{remark}
\label{rem1}
The GSVD of $\bZ$ can be obtained by performing the standard SVD of the matrix ${\tilde \bZ}=\bN^{1/2}\bZ \bM^{1/2}$, that is a GSVD with metrics $\mathbb{I}_n$ on $\mathbb{R}^n$ and  $\mathbb{I}_p$ on $\mathbb{R}^p$. It gives: 
\begin{equation}
{\tilde \bZ}={\tilde \bU} {\tilde \bLambda}  {\tilde \bV}^\top \label{svd_stand}
\end{equation}
and transformation back to the original scale gives:
\begin{equation}
\bLambda =  {\tilde \bLambda},~~
\bU =  \bN^{-1/2}{\tilde \bU},~~
\bV =  \bM^{-1/2}{\tilde \bV}.
\label{back}
\end{equation}
\end{remark}

\paragraph{Principal Components.} The $n$ rows of $\bZ$  are projected with respect to the inner product matrix $\bM$ onto the axes spanned by the vectors $\bv_1,\ldots,\bv_r$ of $\mathbb{R}^p$ (columns of $\bV$) found by solving the sequence (indexed by $i$) of optimization problems:
\begin{equation}
\begin{array}{lll}
 \mbox{maximize } & \| \bZ \bM \bv_i \|^2_{\bN} & \\
\mbox{subject to  } & \bv_i^\top \bM \bv_j=0 & \forall 1 \leq j < i, \\
&  \bv_i^\top \bM \bv_i=1.
\end{array}
\label{pbopt1}
\end{equation}
The solutions $\bv_1,\ldots,\bv_r$ are the eigenvectors of $\bZ^\top\bN\bZ\bM$,  i.e., the  right-singular vectors in  (\ref{gsvd}).  

The principal component scores (also called factor coordinates of the rows hereafter) are the coordinates of the projections of the $n$ rows onto these $r$ axes. Let $\bF$ denote the $n \times r$ matrix of the factor coordinates of the rows. By definition
\begin{equation}
\bF=\bZ \bM\bV, \label{proj_indiv}
\end{equation}

and we deduce from (\ref{gsvd}) that:
\begin{equation}\label{coord_indiv}
\bF=\bU \Lambda.
\end{equation}

Let $\bff_i= \bZ \bM \bv_i$ denote  a column of $\bF$.   The vector $\bff_i \in \mathbb{R}^n$ is called the $i$th principal component (PC) and the solution of (\ref{pbopt1}) gives $\| \bff_i \|^2_{\bN}=\lambda_i $.

\paragraph{Loadings.} The $p$ columns of $\bZ$ are projected with respect to the inner product matrix $\bN$ onto the axes spanned by the vectors $\bu_1,\ldots,\bu_r$ of $\mathbb{R}^n$ (columns of $\bU$) found  by solving the sequence (indexed by $i$) of optimization problems:
\begin{equation}
\begin{array}{lll}
 \mbox{maximize } & \| \bZ^\top \bN \bu_i \|^2_{\bM} & \\
\mbox{subject to  } & \bu_i^\top \bN \bu_j=0 & \forall 1 \leq j < i, \\
&  \bu_i^\top \bN \bu_i=1.
\end{array}
\label{pbopt2}
\end{equation}
The solutions $\bu_1,\ldots,\bu_r$ are the eigenvectors of $\bZ\bM\bZ^\top\bN$, i.e., the left-singular vectors  in  (\ref{gsvd}). 

The loadings (also called factor coordinates of the columns hereafter) are the coordinates of the projections of the $p$ columns onto these $r$ axes.  Let $\bA$ denote the $p \times r$ matrix of the factor coordinates of the columns. By definition
\begin{equation}
\bA=\bZ^\top\bN\bU, \label{proj_var}
\end{equation}
and we deduce from (\ref{gsvd}) that:
\begin{equation}\label{coord_var}
\bA= \bV \Lambda. 
\end{equation}

Let us denote $\ba_i= \bZ^\top \bN \bu_i$ a column of $\bA$. The vector $\ba_i  \in \mathbb{R}^p$  is called the $i$th loadings vector and the solution of (\ref{pbopt2}) gives $\| \ba_i \|^2_{\bM}=\lambda_i$.

\begin{remark}
\label{rem_voca}
The previous definitions give the following entries of  $\bZ=\bU \bLambda  \bV^\top$:
\begin{itemize}
\item The first writing is $\bZ=\bU \bA^\top$ where 
\begin{itemize}
\item[-] $\bU$ is the matrix of the standardized principal components ($\bU=\bF\bLambda^{-1}$),
\item[-] $\bA=\bV\bLambda$ is then the matrix of the loadings of the standardized principal components. 
\end{itemize}
\item The second writing is $\bZ=\bF \bV^\top$ where
\begin{itemize}
\item[-] $\bF$ is the matrix of the principal components,
\item[-] $\bV$ is then the matrix of the loadings of the principal components. 
\end{itemize}

\end{itemize}

\end{remark}

\paragraph{Reduced rank matrix approximation.} PCA  is often seen as a method of least square approximation of the matrix $\bZ$ by a matrix  $\hat{\bZ}$ of rank $q <r$. This  optimal low-rank approximation is defined by:
 
\begin{equation} 
\label{eq_svd_q}
\hat{\bZ} =\bU_q \bLambda_q \bV_q^\top
\end{equation} 

\noindent where $\bU_q$ (resp. $\bV_q$) denote the matrix of the first  $q$ columns of $\bU$ (resp.  $\bV$) and $\bLambda$ is the diagonal matrix of the first $q$ singular values.

The reconstruction matrix  $\hat{\bZ}$ is said to be optimal for matrices of rank $q$ because it satisfies the following condition :
\begin{equation}
\label{opti_qual}
\| \bZ - \hat{\bZ} \|_{\bN,\bM}^2  = \min_{\bX} \| \bZ - \bX \|_{\bN,\bM} ^2
\end{equation}
where $\| \bX \|_{\bN,\bM}^2 = \tr(\bN \bX \bM \bX^\top)$ is a generalization of the  Froebonius norm to the context of generalized SVD with metrics $\bN$ and $\bM$ \citep{abdi2007singular}.  Moreover it can be shown that:
\begin{equation}
\label{opti_reconst}
\| \bZ - \hat{\bZ} \|_{\bN,\bM}^2 =\sum_{i=q+1}^r \lambda_i
\end{equation}

\noindent and accordingly the quality of the reconstruction defined by 
\begin{equation}
\label{pcinert}
\frac{\lambda_1+\ldots+\lambda_q}{\sum_{i=1}^r \lambda_i} 
\end{equation}
is maximized. The quantity (\ref{pcinert}) is interpreted as the proportion of the variance of the data explained by the principal components or as the reconstructed proportion.

\subsection{Standard PCA and standard MCA} \label{sec:standard}

This section presents how standard PCA (for numerical data) and standard MCA (for categorical data) can be obtained from the GSVD of specific matrices $\bZ$, $\bN$, $\bM$. In both cases, the numerical matrix $\bZ$ is obtained by pre-processing the original data matrix $\bX$  and the matrix $\bN$ (resp. $\bM$) is the diagonal matrix of the weights of the rows (resp. the columns) of  $\bZ$.

\paragraph{Standard PCA.}  The data table to be analyzed by PCA comprises $n$ observations described by
$p$ numerical variables, and is represented by the $n \times p$ quantitative matrix $\bX$.  In the pre-processing step, the columns of  $\bX$ are centered and normalized to construct the standardized matrix $\bZ$ (defined such that $\frac{1}{n}\bZ^\top \bZ$ is the linear correlation matrix). The $n$ rows (observations) are usually weighted by $\frac{1}{n}$ and the $p$ columns (variables) are weighted by 1.  It gives $\bN=\frac{1}{n}\mathbb{I}_n$ and $\bM=\mathbb{I}_p$.  The metric $\bM$ indicates that the distance between two observations is the standard euclidean distance between two rows of $\bZ$. The total inertia of $\bZ$  is then equal to $p$. The matrix $\bF$ of the factor coordinates of the observations (principal components) and the matrix $\bA$ of the factor coordinates of the variables (loadings) are calculated directly from (\ref{coord_indiv}) and (\ref{coord_var}). The well-known properties of PCA are the following:
\begin{itemize}
\item[-] Each loading $a_{ji}$ (element of $\bA$) is the linear correlation between the numerical variable $\bx_j$ (the $j$th column of $\bX$) and the $i$th principal component $\bff_i$ (the $i$th column of $\bF$): 
\begin{equation}
a_{ji}=\bz_j^\top\bN\bu_i=r(\bx_j,\bff_i), \label{eqcor}
\end{equation}
where $\bu_i=\frac{\bff_i}{\sqrt{\lambda_i}}$ is the $i$th standardized principal component and $\bz_j$ (resp.  $\bx_j$ ) is the $j$th column of $\bZ$ (resp. $\bX$). 
 \item[-] Each eigenvalue $\lambda_i$ is the variance of the $i$th principal component:  
 \begin{equation}
 \lambda_i=\| \bff_i \|^2_{\bN}=\mbox{Var}( \bff_i) .
 \end{equation}
  \item[-] Each eigenvalue $\lambda_i$ is also the sum of the squared correlations between the $p$ numerical variables and the $i$th principal component:   
\begin{equation}
\lambda_i=\| \ba_i \|^2_{\bM}=\sum_{j=1}^p r^2(\bx_j,\bff_i) .
\end{equation}
\item[-] The contribution of the variable $\bx_j$ to the variance of the  $i$th principal component interprets as a squared loading i.e. squared correlation here:
\begin{equation}
\label{contribquanti}
c_{ji}=a_{ji}^2= r^2(\bx_j,\bff_i).
\end{equation}
\item[-] The total variance of the data matrix $\bZ$ is equal to $p$. The proportion of variance explained by the $i$th principal component is then:
$$ \frac{\lambda_i}{p}.$$
\end{itemize}

\paragraph{Standard MCA.}  The data table to be analyzed by MCA comprises $n$ observations described by
$p$ categorical variables and it is represented by the $n \times p$ matrix $\bX$. Each categorical variable has $m_j$ levels and the sum of the $m_j$'s is equal to $m$. In the pre-processing step, each level is coded as a binary variable and the $n \times m$ indicator matrix $\bG$ is constructed.  Usually MCA  is  performed by applying standard Correspondence Analysis (CA) to this indicator matrix. Here, we provide different ways to calculate the factor coordinates of MCA by applying  a single PCA with metrics to the indicator matrix  $\bG$.

Let $\bZ$ now denote the centered indicator matrix $\bG$.  The $n$ rows (observations) are usually weighted by  $\frac{1}{n}$ and the $m$ columns (levels) are weighted by $\frac{n}{n_s}$, the inverse of the frequency of the level $s$, where  $n_s$ denotes the number of observations that belong to the $s$th level. It gives $\bN=\frac{1}{n}\mathbb{I}_n$ and $\bM=\mbox{diag}(\frac{n}{n_s},s=1\ldots,m)$.  This metric $\bM$ indicates that the distance between two observations is a weighted euclidean distance similar to the $\chi^2$ distance in CA. This distance  gives more importance to rare levels. The total inertia of $\bZ$  with this distance and the weights $\frac{1}{n}$ is equal to $m-p$.  The GSVD of $\bZ$ with these metrics allow a direct calculation using (\ref{coord_indiv}) the matrix $\bF$ of the factor coordinates  of the observations (the principal components).The factor coordinates of the levels however are not obtained directly from the loading matrix $\bA$ defined in (\ref{coord_var}) but from:
 
\begin{equation}
       \bA^* =\bM\bA. \label{eqacm}
\end{equation}
to get back the barycentric property recalled in (\ref{baryprop}) which is central to the interpretation of the results in MCA. The usual properties in MCA are:
\begin{itemize}
\item[-]  Each coordinate $a^*_{si}$  (element of $\bA^*$) is the mean value of the (standardized) factor coordinates of  the observations that belong to level $s$:
\begin{equation}
\label{baryprop}
a^*_{si}=\frac{n}{n_s}a_{si}=\frac{n}{n_s}\bz_s^\top\bN\bu_i=\bar{u}_i^s,
\end{equation}
where $\bz_s$ is the $s$th column of $\bZ$,  $\bu_i=\frac{\bff_i}{\sqrt{\lambda_i}}$ is the $i$th standardized principal component  and $\bar{u}_i^s$ is the mean value of the coordinates of $\bu_i$ associated with the observations that belong to  level $s$. 
 \item[-] Each eigenvalue $\lambda_i$ is the variance of the $i$th principal component:  
\begin{equation}
 \lambda_i=\| \bff_i \|^2_{\bN}=\mbox{Var}( \bff_i) .
\end{equation}
  \item[-]  Each eigenvalue $\lambda_i$ is  the sum of the correlation ratios between the $p$ categorical variables and the $i$th principal component (which is numerical):   
\begin{equation}
\lambda_i=\| \ba_i \|^2_{\bM}=\| \ba_i^* \|^2_{\bM^{-1}}=\sum_{j=1}^p  \eta^2( \bff_i|\bx_j).
\end{equation}
where:
\begin{equation}
\label{defeta2}
 \eta^2( \bff_i|\bx_j)=\frac{\sum_{s \in I_j} \frac{n_s}{n}(\bar{\bff}_i^s-\bar{\bff}_i)^2}{\mbox{Var}( \bff_i)}
 \end{equation} 
where $ I_j$ is the set of indices of the levels of the categorical variable $j$ and $\bar{\bff}_i^s$ is the mean value of the coordinates of $\bff_i$ associated with the observations that belong to  level $s$. Here $\bar{\bff}_i=0$ because the principal components $\bff_i$ are all centered as linear combinaisons of the centered columms of $ \bZ$.

 The correlation ratio $\eta^2( \bff_i|\bx_j)$ measures the link between  the categorical variable $\bx_j$ and the numerical principal component  $\bff_i$ and interprets as the part of the variance of  $\bff_i$  explained by $\bx_j$.

 \item[-] The contribution of the variable $\bx_j$ to the variance of the  $i$th principal component is:
\begin{equation}
\label{contribquali}
 c_{ji}=\sum_{s \in I_j} \frac{n}{n_s} a_{si}^2=\sum_{s \in I_j} \frac{n_s}{n} a_{si}^{*2}=\eta^2( \bff_i|\bx_j)
 \end{equation}
The contribution $c_{ij}$ in (\ref{contribquali}) is also called  a squared loading to mimic PCA where squared loadings are squared correlations (see equation (\ref{contribquanti})).
  
\item[-] The total variance of $\bZ$ is equal to $m-p$. The proportion of variance explained by the $i$th principal component is then:
$$ \frac{\lambda_i}{m-p}.$$

\end{itemize}

\begin{remark} Compared to standard MCA method where correspondence analysis  (CA) is applied to the indicator matrix, we can note that:
\begin{itemize}
\item[-] the total inertia of $\bZ$ (based on the metrics $\bM$ and $\bN$) is equal to  $m-p$, whereas the total inertia in standard MCA is multiplied by $p$ and is equal to $p(m-p)$. This property will be useful to define PCA for mixed data right after.  It will allow to balance the inertia of the numerical data (equal to the number of numerical variables) and the inertia of the categorical data (equal now to the number of levels minus the number of categorical variables),
\item[-] the factor coordinates of the levels are the same. However, the eigenvalues are multiplied by $p$ and factor coordinates of the observations are then  multiplied by $\sqrt{p}$. This property has no impact since results are identical to within one multiplier coefficient.
\end{itemize}
\end{remark}

\section{PCA of a mixture of numerical and categorical data} \label{pcamix_section}

Principal Component Analysis (PCA) methods dealing with a mixture of numerical and categorical variables already exist and have been implemented in functions  like \code{FAMD} of the package \pkg{FactoMineR} or \code{dudi.mix} of the package \pkg{ade4}. 
In the  \proglang{R} package  \pkg{PCAmixdata}, the function  \code{PCAmix}  implements an algorithm presented hereafter as a single PCA with metrics, i.e., based on a Generalized Singular Value Decomposition (GSVD) of pre-processed data. This algorithm includes naturally standard PCA and standard MCA as special cases. Note that \code{FAMD}, \code{dudi.mix} and \code{PCAmix} are three different implementations that give identical results (sometimes up to a constant factor) \footnote{https://chavent.github.io/PCAmixdata/PCAmixcompare.html}.

\subsection{The \code{PCAmix}  algorithm} \label{pcamix_subsection}

The data table to be analyzed by \code{PCAmix} comprises $n$ observations described by $p_1$ numerical variables and $p_2$ categorical variables. It is represented by the $n \times p_1$  numerical data matrix  $\bX_1$ and the $n \times p_2$  categorical data matrix $\bX_2$. Let $m$ denote  the total number of levels of the $p_2$ categorical variables. The \code{PCAmix}  algorithm merges PCA and MCA thanks to the general framework given in Section  \ref{section_gsvd} .  The two steps of \code{PCAmix}  (pre-processing and factor coordinates processing) mimic this general framework with the numerical data matrix  $\bX_1$ and the categorical data matrix $\bX_2$ as inputs. 

\paragraph{Step 1: pre-processing.}
\begin{enumerate}
\item Build the real matrix $\bZ=[\bZ_1,\bZ_2]$ of dimension $n \times (p_1+m) $ where:
\begin{itemize}
\item[$\hookrightarrow$] $\bZ_1$ is the standardized version of $\bX_1$ (as in standard PCA),
\item [$\hookrightarrow$]$\bZ_2$ is  the centered  version of the indicator matrix $\bG$ of $\bX_2$ (as in standard MCA).
\end{itemize}
\item Build the diagonal matrix $\bN$ of the weights of the rows of $\bZ$. The $n$ rows are often weighted by $\frac{1}{n}$, such that $\bN=\frac{1}{n}\mathbb{I}_n$.
\item Build the diagonal matrix $\bM$ of the weights of the columns of $\bZ$:
\begin{itemize}
\item[$\hookrightarrow$] The first $ p_1$  columns (corresponding to the numerical variables)  are weighted by 1 (as in standard PCA).
\item[$\hookrightarrow$] The last $m$  columns (corresponding to the levels of the categorical variables) are weighted by $ \frac{n}{n_s}$  (as in standard MCA), where $n_s,s=1,\dots,m$ denotes the number of observations that belong to the $s$th level. 
\end{itemize}
\end{enumerate}
The metric 
\begin{equation}
\bM=\mbox{diag}(1,\dots,1,\frac{n}{n_1},\dots,\frac{n}{n_m})
\label{metM}
\end{equation}
 indicates that the distance between two rows of $\bZ$ is a mixture of the simple euclidean distance used in PCA (for the  first $p_1$  columns) and the weighted  distance in the spirit of the $\chi^2$ distance used in MCA (for the last $m$   columns). The total inertia of $\bZ$  with this distance and the weights $\frac{1}{n}$ is   equal to $p_1+m-p_2$.

\paragraph{Step 2: factor coordinates processing.}
\begin{enumerate}
\item The GSVD of  $\bZ$  with metrics $\bN$ and $\bM$ gives the decomposition:
$$\bZ=\bU\bLambda\bV^\top$$ 
as defined in (\ref{gsvd}). Let $r$ denote the rank of $\bZ$.
\item The matrix of dimension $n \times r$ of the factor coordinates of the $n$ observations is:
\begin{equation}
\bF=\bZ\bM\bV, 
\end{equation}
or directly computed from the GSVD decomposition as:
\begin{equation}
\bF=\bU \bLambda.
\end{equation}
The columns $\bff_i$ of the matrix $\bF$ are the principal components and the columns $\bu_i=\frac{\bff_i}{\sqrt{\lambda_i}}$ of the matrix $\bU$ are the standardized principal components.
\item   The matrix of dimension $(p_1+m) \times r$ of the factor coordinates of the $p_1$ quantitative variables and the $m$ levels of the $p_2$ categorical variables is:
\begin{equation}
\bA^*= \bM \bV \bLambda =\bM \bA,
\label{astar}
\end{equation}
where $\bA=\bV \bLambda$ is the matrix of the loadings of the standardized principal components (see Remark \ref{rem_voca}).

The matrix $\bA^*$ of factor coordinates splits as follows:
$
\bA^*=
\left[
\begin{array}{c} 
\bA_1^* \\
\bA_2^*  \\
\end{array}
\right]
\begin{array}{l} 
\left.\right\} p_1\\
\left.\right\} m \\
\end{array}
$
where
\begin{itemize}
\item[$\hookrightarrow$]  $\bA^*_1$ contains the factor coordinates of the $p_1$ numerical variables,
\item[$\hookrightarrow$]  $\bA_2^*$ contains the factor coordinates of the $m$ levels.
\end{itemize}
The matrix $\bA^*$ differs from the matrix $\bA$ of the loadings so that $\bA_2^*$  verifies the MCA's barycentic property (\ref{bary}).
\end{enumerate}

\subsection{Properties of PCAmix}
\label{prop_pcamix}
The PCAmix procedure  shares and generalizes the properties of PCA and MCA:

\begin{itemize}
\item[-] For $j=1,\ldots, p_1$:
\begin{equation}
a^*_{ji}=a_{ji}=r(\bx_j,\bff_i), 
\end{equation}
The factor coordinates  in $\bA^*_1$ give the correlations between the $p_1$ quantitative variables and the principal components.

\item[-] For $s=p_1+1, \ldots , p_1+m$:
\begin{equation}
a^*_{si}=\bar{u}_i^s,
\label{bary}
\end{equation}
The factor coordinates  of the $m$ levels in $\bA^*_2$ give the mean values of the standardized principal components  $\bu_i$ for the observations that belong to level $s$.
 
 \item[-] Each eigenvalue $\lambda_i$ is the variance of the $i$th principal component:  
\begin{equation}
\label{maxdisp}
 \lambda_i=\| \bff_i \|^2_{\bN}=\mbox{Var}( \bff_i) .
\end{equation} 
\item[-] Each eigenvalue is the sum of squared correlations (resp. the correlation ratios) between the $p_1$ numerical (resp. $p_2$ categorical) variables and the $i$th principal component (which is numerical):

\begin{align}
\label{maxlink}
\lambda_i &= \| \ba_i \|^2_{\bM}=\| \ba_i^* \|^2_{\bM^{-1}}, \nonumber\\
 &=  \sum_{j=1}^{p_1} a^{2}_{ji} +  \sum_{j=p_1+1}^{p_2} \sum_{s \in I_j} \frac{n}{n_s} a^{2}_{si}, \nonumber \\
  &=  \sum_{j=1}^{p_1} r^2(\bx_j,\bff_i) +  \sum_{j=p_1+1}^{p_2} \eta^2( \bff_i|\bx_j).
 \end{align}
 where $ \eta^2( \bff_i|\bx_j)$ is the correlation ratio defined in (\ref{defeta2}).
 \item[-] The contribution of a variable $\bx_j$ to the variance of the $i$th principal component  is:
 \begin{equation}
\left\{
\begin{array}{ll}
c_{ji}=a_{ji}^2 =r^2(\bx_j,\bff_i)& \mbox{ if the variable }\bx_ j   \mbox{  is numerical},\\
c_{ji}=\sum_{s \in I_j} \frac{n}{n_s} a_{si}^2=\eta^2( \bff_i|\bx_j) & \mbox{ if the variable  } \bx_ j  \mbox{ is categorical},
\end{array}
\right. \label{eq:contrib1}
\end{equation}
The contributions are called \textbf{squared loadings}. Note that the term squared loadings for categorical variables draws an analogy with squared loadings in PCA. Squared loadings are defined as squared correlations for numerical variables and correlation ratios for categorical variables.  
\item[-] The total variance of $\bZ$ is equal to $p_1+m-p_2$. The proportion of variance explained by the $i$th principal component is then:
$$ \frac{\lambda_i}{p_1+m-p_2}.$$
\end{itemize}  

\begin{remark}
\code{PCAmix} computes $q \leq r$ new numerical variables (the principal components) that will ``explain'' or ``extract''  the largest part of the inertia of the  matrix $\bZ$ built from the original data tables $\bX_1$ and $\bX_2$. The  principal components (columns of $\bF$) are  by construction non correlated linear combinations of the columns of $\bZ$ and can be viewed as new synthetic numerical variables with  maximum dispersion (\ref{maxdisp}) and maximum link with the original variables (\ref{maxlink}). 
\end{remark}

\subsection{Graphical outputs of \code{PCAmix}} \label{plotpcamix}
The function \code{plot.PCAmix}  plots the observations, the numerical variables and the  levels  of the categorical variables according to their factor coordinates.

\paragraph{Correlation circle.}
The  map of the quantitative variables, called the correlation circle, gives an idea of the pattern of linear links between the  quantitative variables.  If two columns $\bz_j$ and $\bz_{j'}$ of $\bZ_1$ corresponding to two quantitative variables $\bx_j$ and $\bx_{j'}$ (two columns of $\bX_1$) are well projected on the map i.e. with squared cosine close to 1), the cosine of their angle in projection gives an idea of their correlation in $\R^{n}$ defined  by
$$r(\bx_j,\bx_{j'})=\bz_j^\top\bN\bz_{j'}$$
with  $\bN=\frac{1}{n} \I_n$ in the usual case of observations weighted by $\frac{1}{n}$. 

\paragraph{Levels map.}
The levels map  gives an idea of the pattern of proximities between the levels of (different) categorical variables. 
 If  two levels $\bz_s$ and $\bz_{s'}$ (two columns of  the centered indicator matrix $\bZ_2$)  are well projected on the map, the distance when projected gives an idea of their distance in $\R^{n}$ given by
$$d_\bN^2(\bz_s,\bz_{s'})=(\bz_s-\bz_{s'})^\top\bN(\bz_s-\bz_{s'})$$
which can be interpreted as 1 minus the proportion of observations having both levels $s$ and $s'$. With this distance two levels are similar if they are owned by the same observations.

\paragraph{Squared loadings plot.}

Another graphical output available in \code{plot.PCAmix} is the plot of the variables (numerical and categorical) according to their squared loadings. The map of all the variables gives an idea of the pattern of links between the variables regardless of their type (quantitative or categorical). More precisely, it is easy to verify that the squared loading $c_{ji}$ defined in (\ref{eq:contrib1})  is equal to:
\begin{itemize}
\item[-]  the  squared correlation $r^2( \bff_i, \bx_j)$  if  the variable $\bx_j$  is  numerical,
\item[-] the correlation ratio $\eta^2( \bff_i|\bx_j)$  if the variable $\bx_j$ is categorical. 
\end{itemize}
Coordinates (between 0 and 1) of the variables on this  plot measure the intensity of the links between variables and  principal components and can  be used to interpret principal component maps.

\paragraph{Observations map.}
 The map of the observations (also called principal component map) gives an idea of the pattern of similarities between the  observations. If two observations $\bz_k$ and $\bz_{k'}$ (two rows of $\bZ$) are well projected on the map, their distance in projection gives an idea of their distance in $\R^{p_1+m}$  defined by
$$d_\bM^2(\bz_k,\bz_{k'})=(\bz_k-\bz_{k'})^\top\bM(\bz_k-\bz_{k'})$$
where $\bM$ is defined in (\ref{metM}). This squared distance can be interpreted  as the squared euclidean distance calculated on the standardized numerical variables plus the squared $\chi^2$ distance calculated on the levels of the categorical variables. 
Moreover the position (left, right, up, bottom) of the observations  on the PC's map can  be interpreted in terms of:
\begin{itemize}
\item[-] numerical variables using the property indicating that coordinates on the correlation circle give correlations with PCs,
\item[-] levels of categorical variables using the property indicating that coordinates on the level map are means of PC scores.
\end{itemize}

\subsection{Prediction of PC scores with \code{predict.PCAmix}}
A function to predict scores for new observations on the principal components can be helpful. For example:
\begin{itemize}
\item[-] projecting new observations onto the principal component map of \code{PCAmix},
\item[-] when the PCs are used as synthetic numerical  variables replacing the original variables (quantitative or categorical) in a predictive model (regression or classification for instance).
\end{itemize}

The $i$th principal component of \code{PCAmix}  can be written as a linear combination of the vectors $\bz_1,\ldots,\bz_{p_1+m}$ (columns  of $\bZ$):
$$\bff_i=\bZ\bM\bv_i=\sum_{\ell=1}^{p_1}v_{\ell i}\bz_\ell + \sum_{\ell=p_1+1}^{p_1+m}\frac{n}{n_\ell}v_{\ell i}\bz_\ell. $$

It is then easy to write $\bff_i$ as a linear combination of the vectors $\bx_1,\ldots,\bx_{p_1+m}$  (columns of $\bX=(\bX_1|\bG)$):
\begin{equation}
\bff_i= \beta_{0i}+\sum_{\ell=1}^{p_1+m}\beta_{\ell i} \bx_\ell,
\label{coefpcamix}
\end{equation}
with the coefficients defined as follows:
\begin{align*}
 \beta_{0i} &= -\sum_{\ell=1}^{p_1}v_{\ell i}\frac{{\bar \bx}_\ell}{\hat{\sigma}_\ell} -\sum_{\ell=p_1+1}^{p_1+m}v_{\ell i},\\
\beta_{\ell i} &= v_{\ell i}\frac{1}{\hat{\sigma}_\ell}, \mbox{ for }  \ell=1,\ldots, p_1,\\
\beta_{\ell i}&= v_{\ell i}\frac{n}{n_\ell}, \mbox{ for } \ell=p_1+1,\ldots, p_1+m,
\end{align*}
where ${\bar \bx}_\ell$ and $\hat{\sigma}_\ell$ are respectively the empirical mean and the standard deviation of the column $\bx_\ell$.

The principal components are thereby written in (\ref{coefpcamix}) as a linear combination of the original numerical variables and of the original indicator vectors of the levels of the categorical variables.
The function \code{predict.PCAmix} uses these coefficients  to predict the scores (coordinates) of new observations on the $q \leq r$ first principal components ($q$ is chosen by the user) of \code{PCAmix}.

\subsection{Illustration of \code{PCAmix}} \label{ex_pcamix}

Let us now illustrate the procedure \code{PCAmix} with the data table  \code{housing} of the dataset \code{gironde}. This data table contains $n=542$ municipalities described on $p_1=3$ numerical variables and $p_2=2$ categorical with a total of $m=4$ levels (see Appendix \ref{datades} for the description of the variables).

\begin{verbatim}
R> library("PCAmixdata")
R> data("gironde")
R> head(gironde$housing)
                   density primaryres   houses owners council
ABZAC               131.70      88.77  inf 90%  64.23  sup 5%
AILLAS               21.21      87.52  sup 90%  77.12  inf 5%
AMBARES-ET-LAGRAVE  531.99      94.90  inf 90%  65.74  sup 5%
AMBES               101.21      93.79  sup 90%  66.54  sup 5%
ANDERNOS-LES-BAINS  551.87      62.14  inf 90%  71.54  inf 5%
ANGLADE              63.82      81.02  sup 90%  80.54  inf 5%
\end{verbatim}

In order to explore the mixed data table \code{housing}, a principal component analysis is performed using the function \code{PCAmix}.

\begin{verbatim}
R> split <- splitmix(gironde$housing)
R> X1 <- split$X.quanti 
R> X2 <- split$X.quali 
R> res.pcamix <- PCAmix(X.quanti = X1, X.quali = X2, rename.level = TRUE, graph = FALSE)
R> res.pcamix$eig
      Eigenvalue Proportion Cumulative
dim 1  2.5268771  50.537541   50.53754
dim 2  1.0692777  21.385553   71.92309
dim 3  0.6303253  12.606505   84.52960
dim 4  0.4230216   8.460432   92.99003
dim 5  0.3504984   7.009968  100.00000
\end{verbatim}

Note that the function \code{splitmix} splits a mixed data matrix into two datasets: one with the numerical variables and one with the categorical variables.

The sum of the eigenvalues  is equal to the total inertia $p_1+m-p_2=5$ and the first two dimensions retrieve  71\% of the total inertia. Let us visualize  on these two dimensions the 4 different plots presented in Section \ref{plotpcamix}.

\begin{verbatim}
R> plot(res.pcamix, choice = "ind", coloring.ind = X2$houses, label = FALSE,
      		posleg = "bottomright", main = "(a) Observations")
R> plot(res.pcamix, choice = "levels", xlim = c(-1.5,2.5), main = "(b) Levels")
R> plot(res.pcamix,choice = "cor", main = "(c) Numerical variables")
R> plot(res.pcamix, choice = "sqload", coloring.var = T, leg = TRUE, 
    		 posleg = "topright", main = "(d) All variables")
\end{verbatim}

\begin{figure}[htb]
\begin{center}
     \includegraphics[width=1\textwidth]{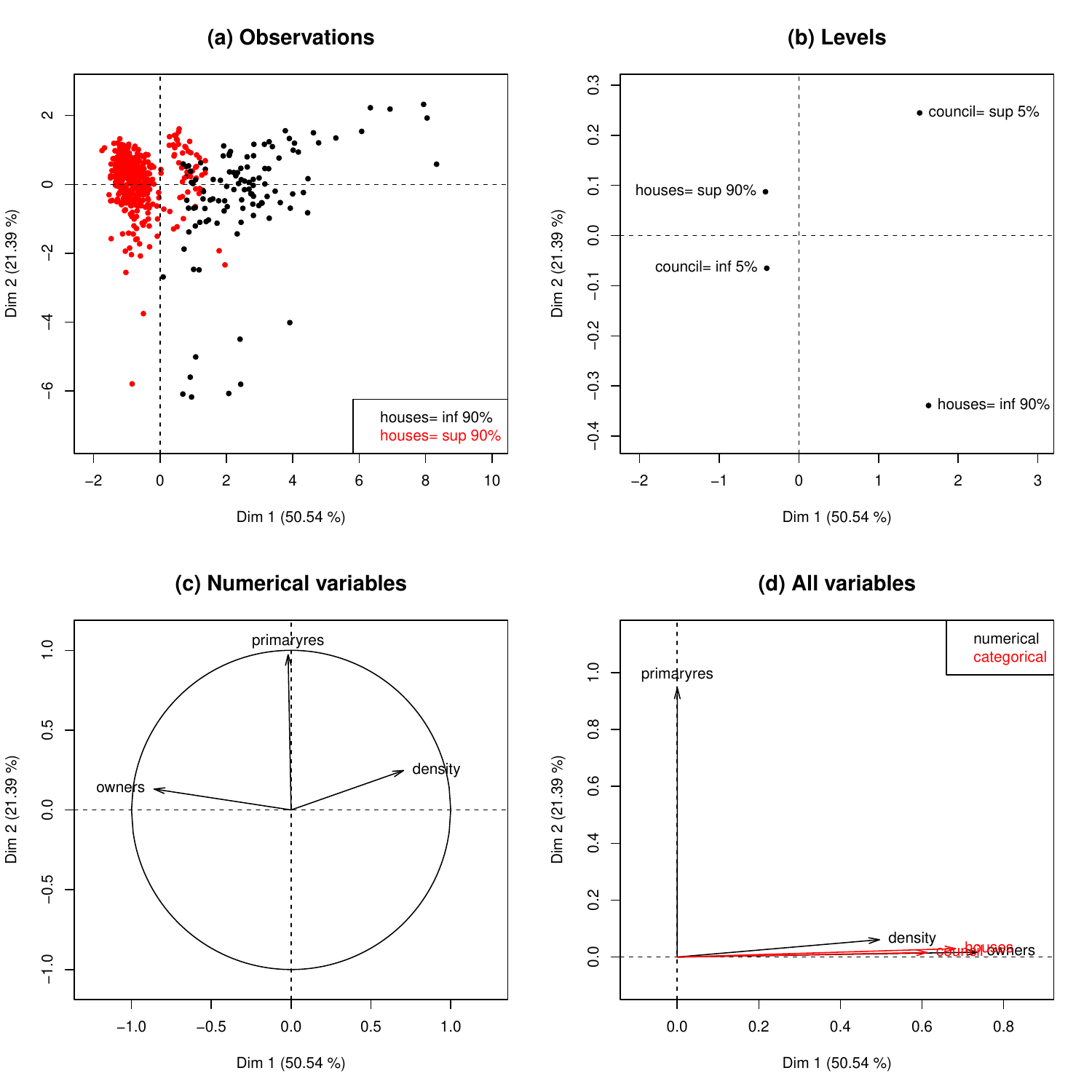}
\caption{Graphical outputs of PCAmix applied to the data table housing}
\label{pcamix_plot}
\end{center}
\end{figure}

Figure~\ref{pcamix_plot}(a) shows the principal component map where the municipalities (the observations) are colored  by their percentage of houses (less than 90\%, more than 90\%). The first dimension (left hand side) highlights municipalities with large proportions of privately-owned properties. The level map in Figure~\ref{pcamix_plot}(b) confirms this interpretation and suggests that municipalities with a high proportion of houses (on the left) have a low percentage of council housing.  The correlation circle in Figure~\ref{pcamix_plot}(c) indicates that population density is negatively correlated with the percentage of home owners and that these two variables discriminate the municipalities on the first dimension.

Figure~\ref{pcamix_plot}(d)  plots  the variables (categorical or numerical) using squared loadings as coordinates.  For numerical variables, squared loadings are squared correlations and for categorical variables squared loadings are correlation ratios. In both cases, they measure the link between the variables and the principal components. It can be observed that the two numerical variables  \code{density} and \code{owners}  and the two categorical variables  \code{houses} and \code{council}  are linked to the first component. On the contrary, the variable \code{primaryres} is clearly orthogonal to these variables and associated with the second component. Note that these links show neither a positive nor a negative association, and the maps Figure~\ref{pcamix_plot}(b) and Figure~\ref{pcamix_plot}(c) are necessary for a more precise interpretation.

In summary, municipalities on the right of the principal component map  have a relatively high proportion of council housing and a small percentage of privately-owned houses, with most accommodation being rented. On the other hand, municipalities on the left hand side are mostly composed of home owners living in their primary residence. The percentage of primary residences also has  a structuring role in the characterization of municipalities in this region of France by defining clearly the second dimension. Indeed the municipalities at the bottom of the map (those with small values on the second dimension) are sea resorts with many secondary residences. For instance the 10 municipalities with the smallest coordinates in the second dimension are well-known resorts on France's Atlantic coast:

\begin{verbatim}
R> sort(res.pcamix$ind$coord[,2])[1:10]
 VENDAYS-MONTALIVET             CARCANS             LACANAU 
          -6.171971           -6.087304           -6.070451 
     SOULAC-SUR-MER GRAYAN-ET-L'HOPITAL     LEGE-CAP-FERRET 
          -5.802359           -5.791642           -5.596315 
     VERDON-SUR-MER             HOURTIN            ARCACHON 
          -5.008545           -4.493259           -4.013374 
              PORGE 
          -3.751233 
\end{verbatim}

\paragraph{Prediction and plot of scores for new observations.}  We will now illustrate how the function \code{predict.PCAmix} can be helpful in predicting the coordinates (scores) of observations not used in \code{PCAmix}.  Here, 100 municipalities are sampled at random  (test set) and the 442  remaining municipalities (training set) are used to perform \code{PCAmix}. The following \proglang{R} code shows how to predict the scores of the municipalities of the test set on the two first PCs obtained with the training set. 

\begin{verbatim}
R> set.seed(10)
R> test <- sample(1:nrow(gironde$housing), 100)
R> train.pcamix <- PCAmix(X1[-test,], X2[-test,], ndim = 2, graph = FALSE)
R> pred <- predict(train.pcamix, X1[test,], X2[test,])
R> head(pred)
                               dim1        dim2
MAZION                   -0.4120140  0.03905247
FLAUJAGUES               -0.6881160 -0.33163728
LATRESNE                  0.7447583  0.65305517
SAINT-CHRISTOLY-DE-BLAYE -0.7006372 -0.33216807
BERSON                   -1.1426625  0.33607088
CHAMADELLE               -1.3781919  0.24609791 
\end{verbatim}

\begin{figure}[htb]
\begin{center}
     \includegraphics[width=0.7\textwidth]{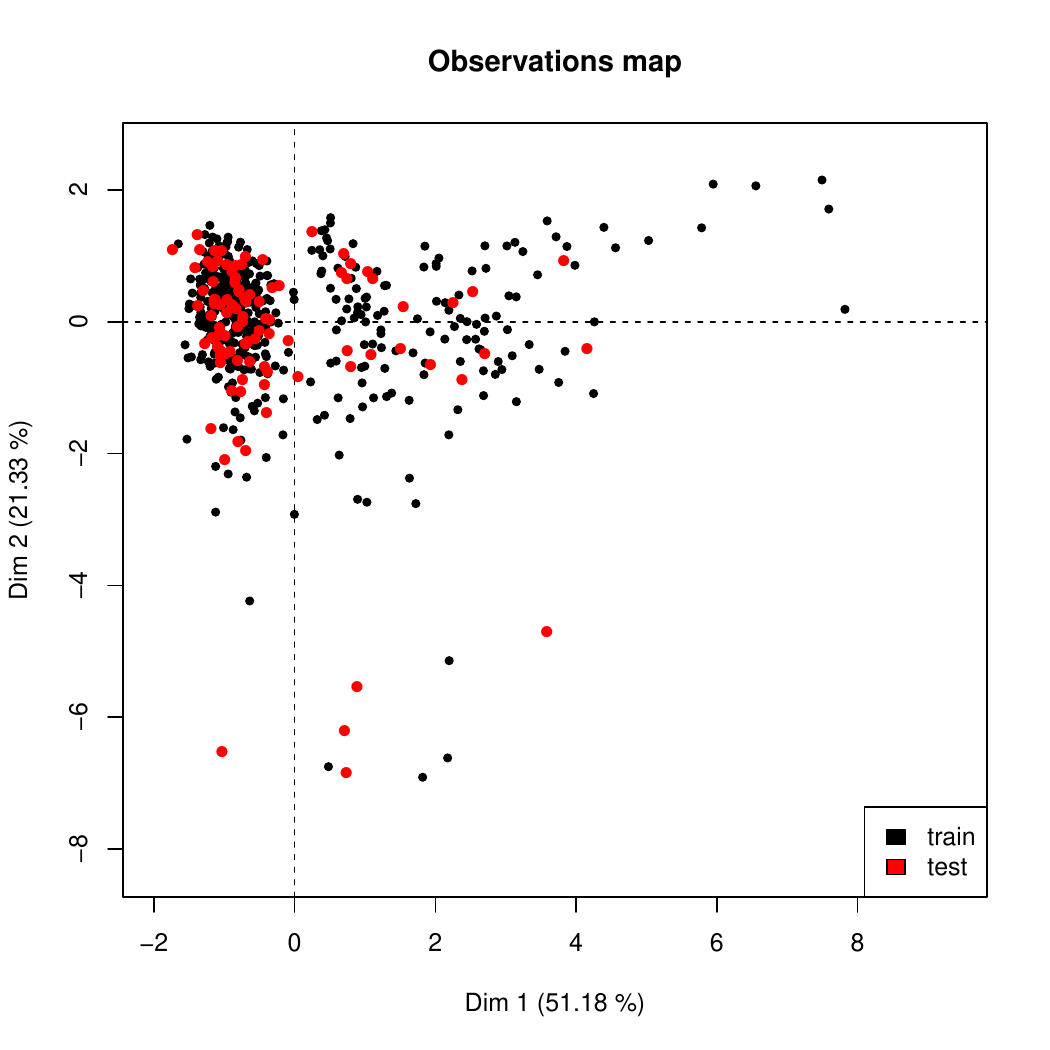}
\caption{Projection of 100 supplementary municipalities (in red) on the PC map of the other 442 municipalities (in black)}
\label{supp_obs_plot}
\end{center}
\end{figure}

These predicted coordinates can be used to plot the 100 supplementary municipalities on the principal component map of the other 442 municipalities (see Figure~\ref{supp_obs_plot}).

\begin{verbatim}
R> plot(train.pcamix, axes = c(1,2), label = FALSE, main = "Observations map")
R> points(pred, col = 2, pch = 16)
R> legend("bottomright", legend = c("train","test"), fill = 1:2, col = 1:2)
\end{verbatim}

\paragraph{Supplementary variables.} The function \code{supvar.PCAmix} calculates the coordinates of  supplementary variables (numerical or categorical) on the maps of \code{PCAmix}. More precisely this function builds an \proglang{R}  object of class  \code{PCAmix} including the supplementary coordinates. 
For instance let us consider the numerical variable \code{building} of the dataset  \code{environment}  and the categorical variable \code{doctor} of the dataset  \code{services}  as supplementary variables (see  Appendix \ref{datades} for description of these two variables). 

\begin{verbatim}
R>  X1sup <- gironde$environment[ , 1, drop = FALSE]
R>  X2sup <- gironde$services[ , 7, drop = FALSE]
R>  res.sup <- supvar(res.pcamix, X1sup, X2sup, rename.level = TRUE)
R>  res.sup$quanti.sup$coord[ , 1:2, drop = FALSE]
              dim1      dim2
building 0.6945295 0.1884711
R>  res.sup$levels.sup$coord[ ,1:2]
                     dim1         dim2
doctor=0      -0.44403187 -0.006224754
doctor=1 to 2  0.07592759 -0.112352412
doctor=3 or +  1.11104073  0.099723319
\end{verbatim}

\begin{figure}[htb]
\begin{center}
     \includegraphics[width=0.95\textwidth]{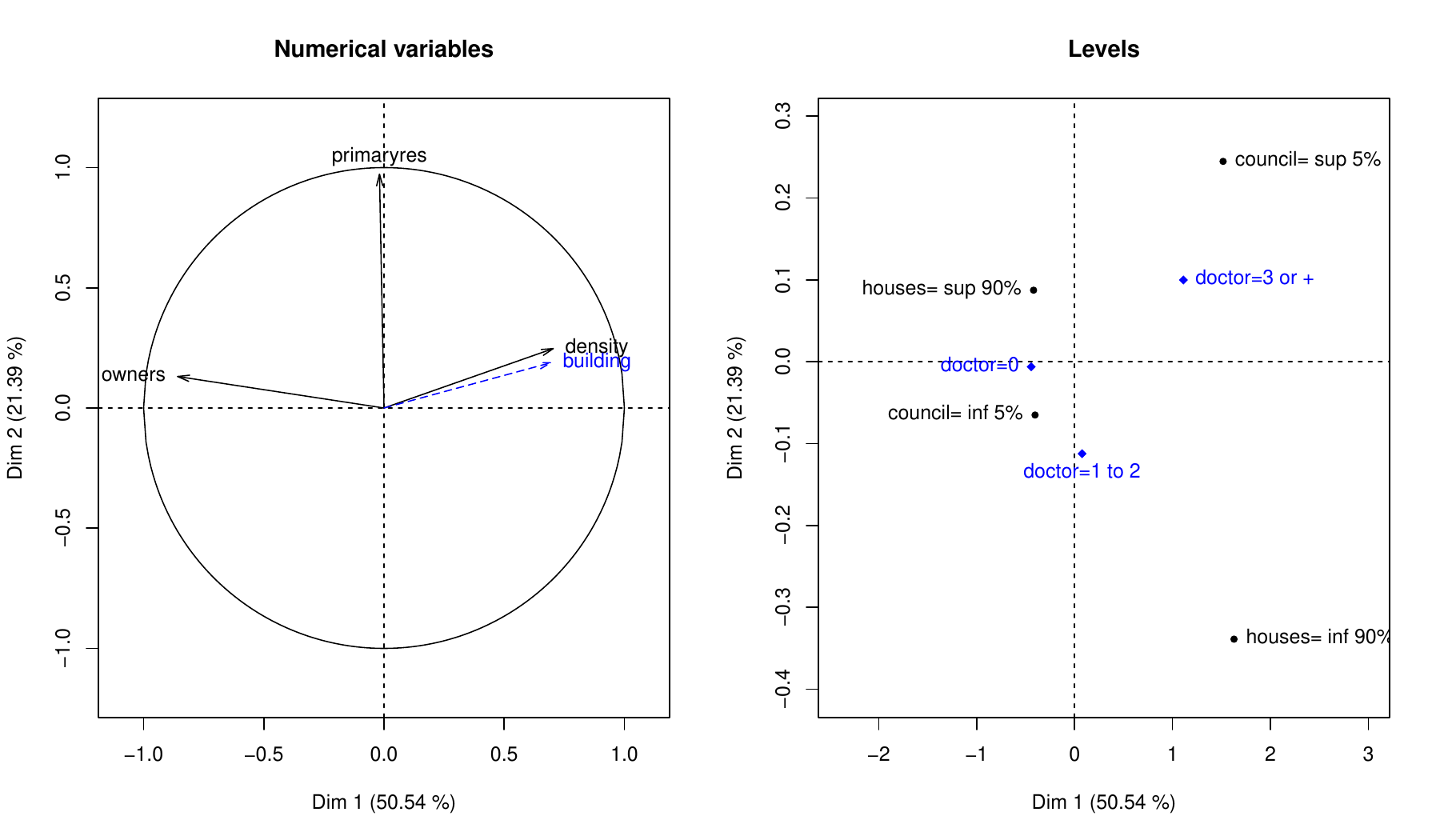}
\caption{In blue, projection of the supplementary numerical variable building (left) and projection of the  levels of the supplementary categorical variable doctor (right)}
\label{supp_var_plot}
\end{center}
\end{figure}

The coordinates of the supplementary numerical variables  \code{building} are still correlations. For instance, the correlation between  \code{building} and the first PC is equal to 0.69. The coordinates of the  levels of the supplementary categorical variables are still mean values.  For instance the coordinate -0.44 of the level \code{doctor=0} is the mean value of the municipalities with 0 doctors on the first standardized PC. They are probably mostly left of the PC map. 
Graphical outputs including these supplementary variables and the original ones can be obtained as previously with the  function \code{plot.PCAmix}, see Figure~\ref{supp_var_plot}.  
\begin{verbatim}
R> plot(res.sup, choice = "cor", main = "Numerical variables")
R> plot(res.sup, choice = "levels", main = "Levels", xlim = c(-2,2.5))
\end{verbatim}

\section{Orthogonal rotation in PCA of mixed data} \label{pcarot_section}

It is common practice in PCA to apply a rotation procedure to loadings to simplify interpretation of the principal components.  The well known varimax rotation procedure \citep{kaiser1958varimax} is implemented in the \proglang{R}  function \code{varimax} of the \pkg{stats} package but this procedure fits only for numerical data. The function \code{PCArot}  of the package \pkg{PCAmixdata} implements a generalization of the varimax procedure to the case of mixed data \citep{chavent2012orthogonal}. The rotation procedure \code{PCArot} applies to the (standardized) principal components of  \code{PCAmix} to get  either large  (close to 1) or small (close to 0) squared loadings. Indeed in \code{PCAmix} the squared loadings  are squared correlations for numerical variables and  correlation ratios for categorical variables measuring then the link between the variables (numerical or categorical) and the principal components. The rotation procedure  \code{PCArot}  is therefore applied to the first $q$ principal components of the procedure \code{PCAmix} where $q$ is chosen by the user.

\subsection{The \code{PCArot} algorithm}

We have seen  that \code{PCAmix} is essentially a GSVD that gives the decomposition:
$$\bZ=\bU\bLambda\bV^\top$$ 
defined in (\ref{gsvd}). The columns of $\bU$ are the standardized principal components (PCs) and the columns of $\bA=\bV \bLambda$ are the loading vectors of the standardized principal components.   The \code{PCArot}  procedure rotates the matrix $\bU_q$  of the  first $q$ standardized PCs and the matrix $\bA_q$ of the first  $q$  loading vectors.  Rotating the loadings $\bA_q$ of the standardized PCs (rather than loadings $\bV_q$ of the PCs) provides rotated scores that continue to be uncorrelated. The \code{PCArot} procedure does so by following \cite{kiers1991simple} proposal to maximize Kaiser's varimax function applied to squared loadings for quantitative variables and `pseudo' squared loadings for categorical variables. Here the \code{PCArot}  procedure uses the alternative algorithm proposed by \cite{chavent2012orthogonal} which expressed the squared loadings for categorical variables somewhat differently, using all elements of $\bA_q$, as follows.

Let $\bT$ be a $q \times q$ orthonormal rotation matrix.  Let $\bU_{\mbox{\tiny rot}}=\bU_q\bT$ denote the matrix of the rotated standardized PCs  and $\bA_{\mbox{\tiny rot}}=\bA_q\bT$ denote the matrix of the rotated loading vectors. In varimax rotation, the matrix $\bT$ is computed by maximizing the variance of the contributions of the variables which interprets as squared loadings. The squared loadings defined in (\ref{eq:contrib1}) write after rotation:

\begin{equation}
\left\{
\begin{array}{ll}
c_{ji,\mbox{\tiny rot}}=a_{ji,\mbox{\tiny rot}}^2 & \mbox{ if the variable }\bx_ j  \mbox{  is numerical},\\
c_{ji,\mbox{\tiny rot}}=\sum_{s \in I_j} \frac{n}{n_s} a_{si,\mbox{\tiny rot}}^2& \mbox{ if the variable  } \bx_ j  \mbox{ is categorical},
\end{array}
\right. \label{eq:contrib1rot}
\end{equation}
They measure the links (squared correlations or correlation ratios) between the principal components after rotation and the variables.  Maximizing the variance of the squared loadings after rotation leads to high values of squared loadings for several variables and low for the remainder while leaving the quality of the matrix reconstruction unchanged. The varimax rotation problem is  then rephrased as 
\begin{equation}
\begin{array}{ll}
\displaystyle \max_{\bT} & \{f(\bT) | \bT \bT^\top=\bT^\top\bT=\I_q\},
\end{array}
 \label{pboptim}
\end{equation}
where
\begin{equation}
f(\bT) =\sum_{i=1}^q\sum_{j=1}^p (c_{ji,\mbox{\tiny rot}})^2 - \frac{1}{p} \sum_{i=1}^q  \left(\sum_{j=1}^p c_{ji,\mbox{\tiny rot}}\right)^2. \label{varipcamix} 
\end{equation}
  The objective function (\ref{varipcamix}) also writes (see Appendix \ref{appendix:proof1}):
\begin{equation}
f(\bT) =\sum_{i=1}^q\sum_{j=1}^p ({\tilde c}_{ji,\mbox{\tiny rot}})^2 - \frac{1}{p} \sum_{i=1}^q  \left(\sum_{j=1}^p {\tilde c}_{ji,\mbox{\tiny rot}}\right)^2. \label{varipcamix2} 
\end{equation}
where ${\tilde c}_{ji,\mbox{\tiny rot}}$ are the squared loadings after rotation obtained with the standard SVD
$$
{\tilde \bZ}={\tilde \bU} {\tilde \bLambda}  {\tilde \bV}^\top ,
$$
of the matrix ${\tilde \bZ}=\bN^{1/2}\bZ \bM^{1/2}$ (see Remark \ref{rem1}). The rotation procedure proposed by  \citep{chavent2012orthogonal}  uses the standard SVD of ${\tilde \bZ}$ to optimize  the objective function (\ref{varipcamix2}). This procedure summarized in Appendix~\ref{Appendix:pcarot} finds an optimal rotation matrix $\bT$ and gives: 
\begin{align}
{\tilde \bU}_{\mbox{\tiny rot}}&={\tilde\bU}_q\bT \\
{\tilde \bA}_{\mbox{\tiny rot}}&={\tilde\bA}_q\bT 
\end{align}

\paragraph{Rotated factor coordinates processing.}
\begin{enumerate}
\item  The matrix of dimension $(p_1+m) \times q$ of the rotated factor coordinates of the $p_1$ quantitative variables and the $m$ levels of the $p_2$ categorical variables is (see \ref{adac}): 
\begin{equation}
\bA_{\mbox{\tiny rot}}^*= \bM \bA_{\mbox{\tiny rot}}=\bM^{1/2}\tbA_{\mbox{\tiny rot}}.
\end{equation}

$\bA_{\mbox{\tiny rot}}^*$ is split as follows:
$
\bA_{\mbox{\tiny rot}}^*=
\left[
\begin{array}{c} 
\bA_{1,\mbox{\tiny rot}}^* \\
\bA_{2,\mbox{\tiny rot}}^*  \\
\end{array}
\right]
\begin{array}{l} 
\left.\right\} p_1\\
\left.\right\} m \\
\end{array}
$
where
\begin{itemize}
\item[$\hookrightarrow$]  $\bA^*_{1,\mbox{\tiny rot}}$ contains the rotated factor coordinates of the $p_1$ numerical variables,
\item[$\hookrightarrow$]  $\bA_{2,\mbox{\tiny rot}}^*$ contains the rotated factor coordinates of the $m$ levels.
\end{itemize}
\item The variance $\lambda_{i,\mbox{\tiny rot}}$ of the $i$th rotated principal component is calculated as:
\begin{equation}
\lambda_{i,\mbox{\tiny rot}}=\| \ba_{i,\mbox{\tiny rot}} \|^2_{\bM}=\| \tba_{i,\mbox{\tiny rot}} \|^2_{\mathbb{I}_{p_1+m}},
\end{equation}
where $\ba_{i,\mbox{\tiny rot}}$ (resp.$\tba_{i,\mbox{\tiny rot}} $)  is the $i$th column of $\bA_{\mbox{\tiny rot}}$ (resp. $\tbA_{\mbox{\tiny rot}}$). 

Let $\Lambda_{\mbox{\tiny rot}}=\mbox{diag}(\sqrt{\lambda_{1,\mbox{\tiny rot}}},\ldots,\sqrt{\lambda_{q,\mbox{\tiny rot}}})$ denote the diagonal matrix of the standard deviations of the $q$ rotated principal components. 
\item The matrix of dimension $n \times q$ of the  rotated factor coordinates of the $n$ observations is:
\begin{equation}
\bF_{\mbox{\tiny rot}}=\bU_{\mbox{\tiny rot}}\Lambda_{\mbox{\tiny rot}}=\bN^{-1/2} \tbU_{\mbox{\tiny rot}}\Lambda_{\mbox{\tiny rot}}. 
\end{equation}
\end{enumerate}

\begin{remark}  For numerical data, \code{PCArot} is the  standard varimax procedure defined by Kaiser (1958) for rotation in PCA. For categorical data,   \code{PCArot} is an orthogonal rotation procedure for Multiple Correspondence Analysis (MCA). 
\end{remark}

\subsection{Properties of \code{PCArot}}

The properties used  to interpret the graphical outputs of \code{PCAmix} remain true after rotation:
\begin{itemize}
\item[-] the rotated factor coordinates of the $p_1$ numerical variables (the  first $p_1$ rows of $\bA_{\mbox{\tiny rot}}^*$) are correlations  with the rotated principal components (the columns of $\bF_{\mbox{\tiny rot}}$),  
\item[-] the rotated factor scores of the $m$ levels  (the $m$ last rows of $\bA_{\mbox{\tiny rot}}^*$) are mean values of the (standardized) rotated factor coordinates of  the observations that belong these levels. 
\end{itemize}

The contribution (squared loading) of the variable $\bx_j$ to the variance of the rotated principal component $\bff_{i,\mbox{\tiny rot}}$ is calculated directly from the matrix $\tbA_{\mbox{\tiny rot}}$ with:

 \begin{equation}
\left\{
\begin{array}{ll}
c_{ji,\mbox{\tiny rot}}=\ta_{ji,\mbox{\tiny rot}}^2=r^2( \bff_{i,\mbox{\tiny rot}},\bx_j) & \mbox{ if the variable } \bx_j  \mbox{  is numerical},\\
c_{ji,\mbox{\tiny rot}}=\sum_{s \in I_j}  \ta_{si,\mbox{\tiny rot}}^2=\eta^2( \bff_{i,\mbox{\tiny rot}}|\bx_j) & \mbox{ if the variable } \bx_j  \mbox{ is categorical}.
\end{array}
\right.
\end{equation}
The squared loadings after rotation are then the squared correlation or correlation ratio between the variables and the rotated principal components.

The function \code{plot.PCAmix} presented in Section \ref{ex_pcamix} plots the observations, the numerical variables and the levels of the categorical variables according to their factor coordinates after rotation. It also plots variables according to their squared loadings after rotation. The interpretation rules given in Section \ref{plotpcamix} remain true.

\subsection{Prediction of rotated PC scores with \code{predict.PCAmix}}

\code{PCArot} computes $q$ new non correlated numerical variables called rotated principal components that  will explain the same part of inertia than \code{PCAmix}  but with simpler interpretation. Let us show that the  rotated principal components (columns of $\bF_{\mbox{\tiny rot}}$) are  linear combination of the columns of $\bZ$. 

First it can be showed (see Appendix~\ref{appendix:proof}) that:
\begin{equation}
\bF_{\mbox{\tiny rot}}=\bZ \bV_{\mbox{\tiny rot}}, \label{defFrot}
 \end{equation}
with 
\begin{equation}
\bV_{\mbox{\tiny rot}}=\bM^{1/2} \tbV_q \tbLambda_q^{-1} \bT \bLambda_{\mbox{\tiny rot}}, \label{defVrot}
 \end{equation}

 and
 \begin{equation}
\bT=\tbU_q^\top \tbU_{\mbox{\tiny rot}}. \label{defT}
 \end{equation}

It follows that the $i$th rotated principal component $\bff_{i,\mbox{\tiny rot}}$ of \code{PCArot} writes as a linear combination of the vectors $\bz_1,\ldots,\bz_{p_1+m}$ (columns  of $\bZ$):
\begin{equation}
\bff_{i,\mbox{\tiny rot}}=\bZ\bv_{i,\mbox{\tiny rot}}=\sum_{\ell=1}^{p_1+m}v_{\ell i,\mbox{\tiny rot}}\bz_\ell. 
 \end{equation}
 
It is then easy to write $\bff_{i,\mbox{\tiny rot}}$ as a linear combination of the vectors $\bx_1,\ldots,\bx_{p_1+m}$  (columns of $\bX=(\bX_1|\bG)$):
\begin{equation}
\bff_{i,\mbox{\tiny rot}}= \beta_{0i,\mbox{\tiny rot}}+\sum_{\ell=1}^{p_1+m}\beta_{\ell i,\mbox{\tiny rot}} \bx_\ell,
\label{coefpcarot}
 \end{equation}

with the coefficients

\begin{align*}
 \beta_{0i,\mbox{\tiny rot}} &= -\sum_{\ell=1}^{p_1}v_{\ell i,\mbox{\tiny rot}}\frac{{\bar \bx}_\ell}{\hat{\sigma}_\ell} -\sum_{\ell=p_1+1}^{p_1+m}v_{\ell i,\mbox{\tiny rot}}\frac{n}{n_\ell} {\bar \bx}_\ell,\\
\beta_{\ell i,\mbox{\tiny rot}} &= v_{\ell i,\mbox{\tiny rot}}\frac{1}{\hat{\sigma}_\ell}, \mbox{ for }  \ell=1,\ldots, p_1,\\
\beta_{\ell i,\mbox{\tiny rot}}&= v_{\ell i,\mbox{\tiny rot}}\frac{n}{n_\ell}, \mbox{ for } \ell=p_1+1,\ldots, p_1+m,
\end{align*}
where ${\bar \bx}_\ell$ and $\hat{\sigma}_\ell$ are respectively the empirical mean and the standard deviation of the column $\bx_\ell$.

The rotated principal components are thereby in (\ref{coefpcarot}) linear combinations of the original numerical variables and of the original indicator vectors of the levels of the categorical variables.
The function \code{predict.PCAmix} uses these coefficients  to predict the scores (coordinates) of new observations on the $q$ rotated principal components of \code{PCArot}.

\subsection{Illustration of \code{PCArot}} \label{ex_pcarot}

Let us now illustrate the procedure \code{PCArot} with the mixed data table  \code{housing} already used in Section\ref{ex_pcamix}.  Let us first create a data frame without the first ten  municipalities (used later for prediction purposes).

\begin{verbatim}
R> library("PCAmixdata")
R> data("gironde")
R> train <- gironde$housing[-c(1:10), ]
R> split <- splitmix(train)
R> X1 <- split$X.quanti 
R> X2 <- split$X.quali 
R> res.pcamix <- PCAmix(X.quanti=X1, X.quali = X2, rename.level = TRUE, graph = FALSE)
R> res.pcamix$eig
      Eigenvalue Proportion Cumulative
dim 1  2.5189342  50.378685   50.37868
dim 2  1.0781913  21.563825   71.94251
dim 3  0.6290897  12.581794   84.52430
dim 4  0.4269180   8.538361   93.06267
dim 5  0.3468667   6.937335  100.00000
\end{verbatim}

The first $q=3$  principal components of \code{PCAmix}  retrieve  84.5\% of the total inertia. In order to improve the interpretation of these 3 components without adversely affecting the proportion of explained inertia  we perform a rotation using the function \code{PCArot}. 

\begin{verbatim}
R> res.pcarot<-PCArot(res.pcamix, dim = 3, graph = FALSE)
R> res.pcarot$eig #variance of the rotated PCs
         Variance Proportion
dim1.rot 1.919546   38.39092
dim2.rot 1.057868   21.15737
dim3.rot 1.248801   24.97601
\end{verbatim}

The spread of the proportion of variance in the three dimensions is modified but the rotated principal components still contain 84.5\% of the total inertia: 
\begin{verbatim}
R> sum(res.pcarot$eig[ ,2])
[1] 84.5243
\end{verbatim}

The rotation also modifies squared loadings with more clear association after rotation  between the third principal component and the variable density. Indeed the squared correlation between \code{density} and the third PC is equal to 0.39 before rotation and increases to 0.9 after rotation. 

\begin{verbatim}
R> res.pcamix$sqload[ ,1:3]
           dim 1 dim 2 dim 3
density     0.49  0.07  0.39
primaryres  0.00  0.94  0.02
owners      0.73  0.02  0.00
houses      0.68  0.03  0.03
council     0.61  0.01  0.18
\end{verbatim}

\begin{verbatim}
R> res.pcarot$sqload
           dim1.rot dim2.rot dim3.rot
density        0.04     0.01     0.90
primaryres     0.00     0.96     0.01
owners         0.48     0.03     0.25
houses         0.63     0.03     0.08
council        0.76     0.03     0.01
\end{verbatim}

Because the rotation improves the interpretation of the third principal component while the second component hardly changed, we plot the observations and the variables on the dimensions 1 and 3.

\begin{verbatim}
R> plot(res.pcamix, choice = "ind", axes = c(1,3), label = FALSE, 
      main = "Observations before rotation")
R> plot(res.pcarot, choice = "ind", axes = c(1,3), label = FALSE, 
      main = "Observations after rotation")
R> plot(res.pcamix, choice = "sqload", axes = c(1,3), 
      main="Variables before rotation", coloring.var = TRUE, leg = TRUE)
R> plot(res.pcarot, choice = "sqload", axes = c(1,3), 
      main="Variables after rotation", coloring.var = TRUE, leg = TRUE)
\end{verbatim}

\begin{figure}[htb]
\begin{center}
\includegraphics[width=1\textwidth]{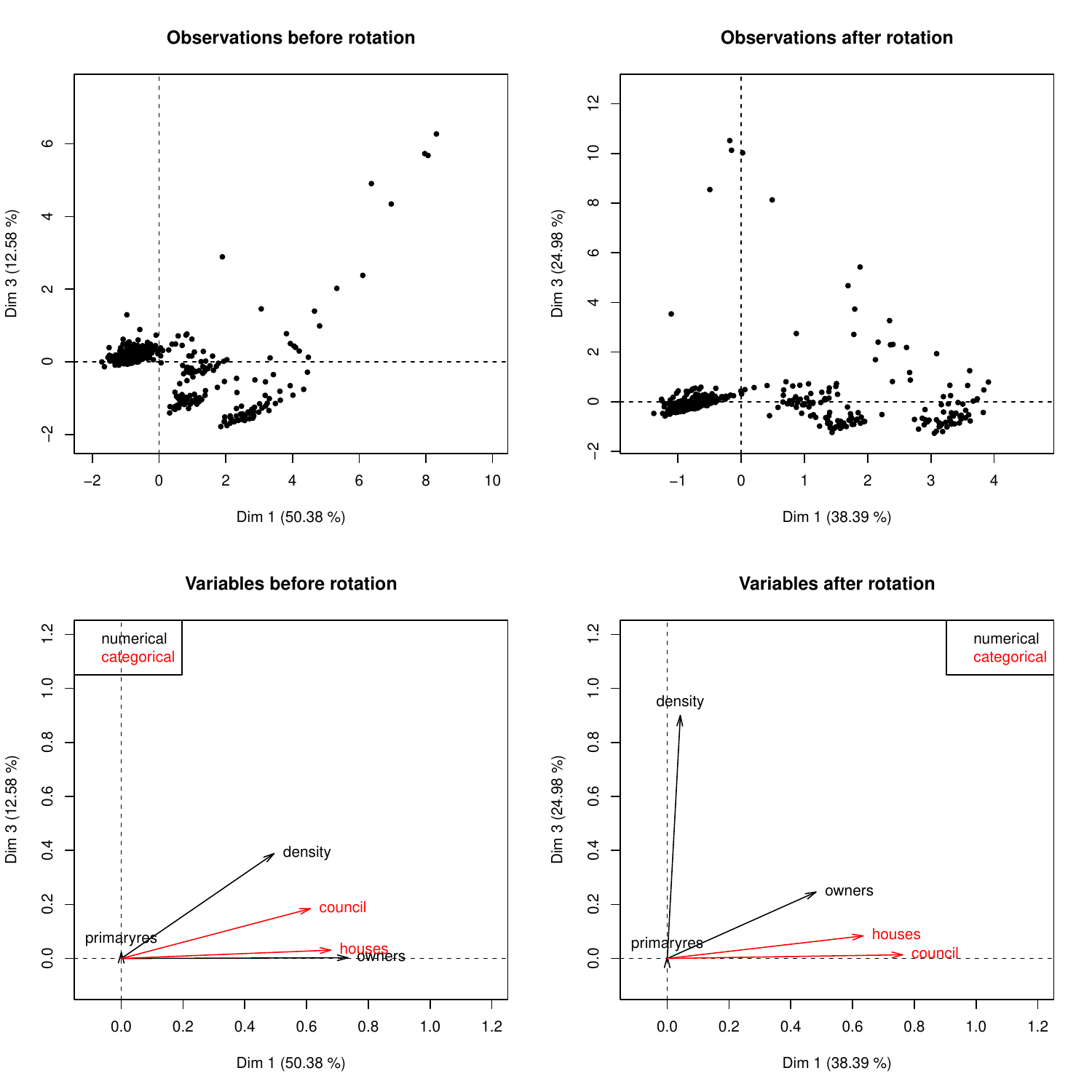} 
\caption{Graphical outputs of PCAmix applied to the data table housing (deprived of the 10 first rows) before rotation (left) and after rotation with PCArot (right).}
\label{pcarot_fig}
\end{center}
\end{figure}

Figure~\ref{pcarot_fig} shows how the variable \code{density} is more clearly linked  after rotation to the third  principal component. Indeed, after rotation, the coordinates of the variable \code{density} on the y-axis is equal to 0.9 (the squared correlation between \code{density} and the 3rd rotated principal component). The municipalities  at the top of the plot of the observations after rotation are then characterized by their population density. Note that the benefit of using rotation on this dataset is limited.

\paragraph{Prediction after rotation.}  Let us now predict the scores of the 10 first municipalities of the data table \code{housing} on the rotated principal components of \code{PCArot}. 
\begin{verbatim}
R> test <- gironde$housing[1:10, ]
R> splitnew <- splitmix(test)
R> X1new <- splitnew$X.quanti
R> X2new<-splitnew$X.quali
R> pred.rot <- predict(object = res.pcarot, X.quanti = X1new, X.quali = X2new)


R> pred.rot
                     dim1.rot   dim2.rot    dim3.rot
ABZAC               3.2685436  0.3494533 -0.85177749
AILLAS             -0.7235629  0.1200285 -0.22254455
AMBARES-ET-LAGRAVE  2.8852451  0.9823515 -0.03451571
AMBES               1.7220716  1.1590890 -0.78227835
ANDERNOS-LES-BAINS  0.3423361 -2.6886415  0.90574890
ANGLADE            -0.9131248 -0.4514258 -0.20108349
ARBANATS           -0.6653760  0.4217893  0.13105217
ARBIS              -0.7668742  0.3099338 -0.23304721
ARCACHON            1.8825083 -4.4533014  2.36935740
ARCINS             -0.6896492  0.2060403 -0.09049882
\end{verbatim}

\begin{figure}[htb]
\begin{center}
\includegraphics[width=0.5\textwidth]{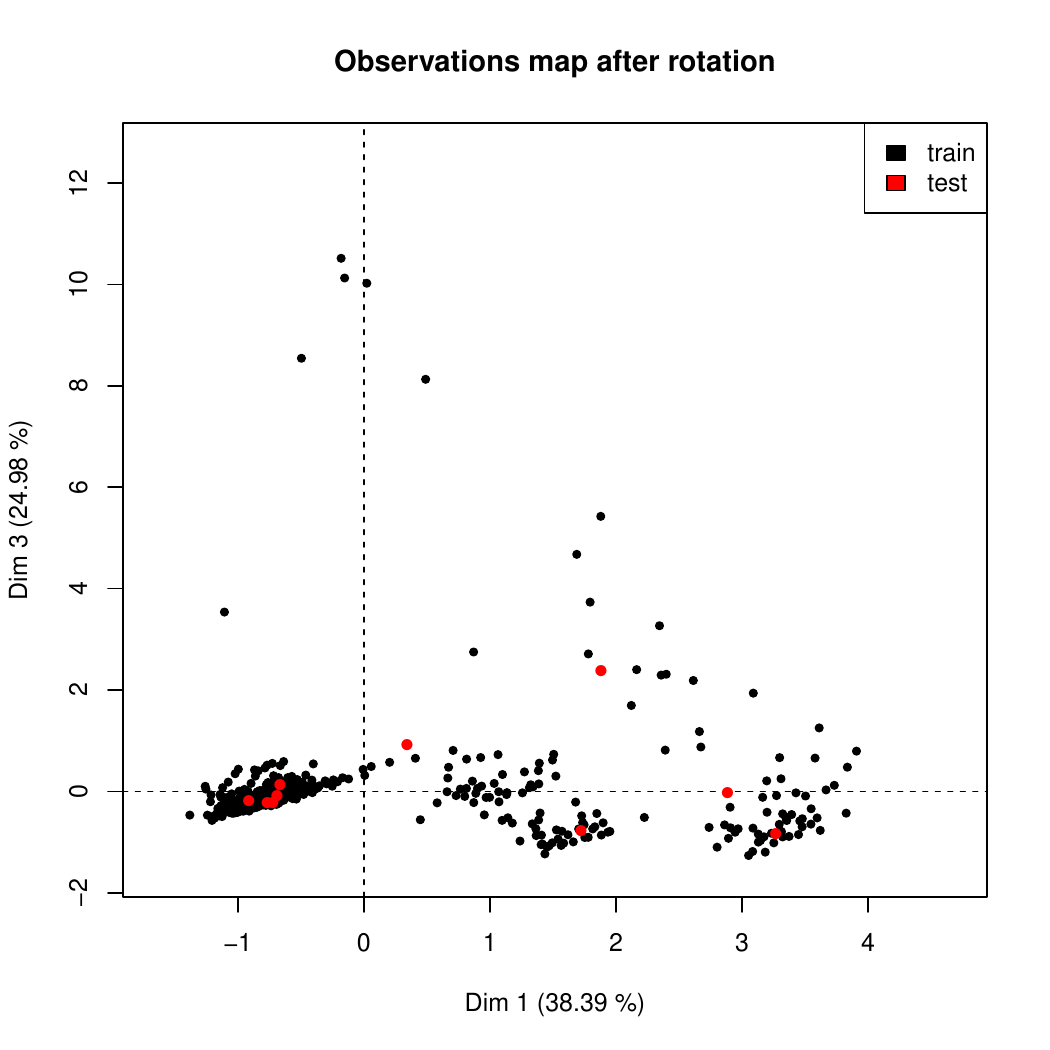} 
\caption{Projection of 10 supplementary municipalities (in red) on the map after rotation.} 
\label{supp_obs_rot}
\end{center}
\end{figure}

These predicted coordinates can be used to plot the 10 supplementary municipalities on the rotated principal component map of the other 532 municipalities (Figure \ref{supp_obs_rot}).

\begin{verbatim}
R> plot(res.pcarot, axes = c(1,3), label = FALSE, main = "Observations map after rotation")
R> points(pred.rot[ ,c(1,3)], col = 2, pch = 16)
R> legend("topright", legend = c("train","test"), fill = 1:2, col = 1:2)
\end{verbatim}

\section{Multiple factor analysis of mixed data} \label{mfamix_section}

Multiple factor analysis \citep{escofier1994multiple,abdi2013multiple} is  a  multivariate analysis method for multi-table data where observations are described by several groups of variables.  The straightforward analysis obtained by concatenating all variables in a single data table  has the drawback of giving more importance to groups with strong structure. The main idea in Multiple Factor Analysis (MFA)  is therefore to give the same importance to each group by weighting each variable by the inverse of the variance of the first principal component of its group. In standard MFA, the nature of the variables (categorical or numerical) can vary from one group to another but the variables within a  group must be of the same nature. The \code{MFAmix} procedure proposed in this paper  works with mixed data even within a group.

\subsection{The \code{MFAmix} algorithm}

Here the $p$ variables  are separated into $G$ groups. The types of variables within a group can be mixed.  Each group  is represented by a data matrix  $\bX^{\g}= [\bX_1^{\g},\bX_2^{\g}]$ where $\bX_1^{\g}$ (resp. $\bX_2^{\g}$) contains the numerical (resp. categorical) variables of group $g=1,\ldots,G$.  The numerical columns (resp. the categorical columns) of the matrices $\bX^{\g}$ are concatenated in a  global numerical data matrix $\bX_1= [\bX_1^{(1)},\ldots,\bX_1^{(G)}]$ (resp.  a global categorical data matrix $\bX_2= [\bX_2^{(1)},\ldots,\bX_2^{(G)}]$). Let $\bZ$ denote the matrix constructed with $\bX_1$ and $\bX_2$ as described  in the pre-processing step of \code{PCAmix} in Section~\ref{pcamix_subsection}. The matrix $\bZ$ has then $n$ rows and $p_1+m$ columns where $p_1=p_1^{(1)}+\ldots+p_1^{(G)}$  and $m=m^{(1)}+\ldots+m^{(G)}$. Each column of $\bZ$ is either a numerical variable  (standardized)  or the  indicator vector of a level (centered).  Let $\bN=\frac{1}{n} \I_n$ and $\bM=\mbox{diag}(1,\dots,1,\frac{n}{n_1},\dots,\frac{n}{n_m})$ be the diagonal matrices of the weights of the rows and columns of $\bZ$.  

The  \code{MFAmix}  algorithm is a  procedure where the first step modifies the weights of the columns of $\bZ$ to equilibrate the importance of the groups in a global \code{PCAmix} analysis.

\paragraph{Step 1: weighting step.}

\begin{enumerate}
\item For $g=1,\ldots, G$, compute the first eigenvalue $\lambda_1^{\g}$  of \code{PCAmix}  applied to  $\bX^{\g}$.
\item Build the diagonal matrix $\bP$ of the weights $\frac{1}{\lambda_1^{(t_k)}}$ where $t_k \in \{1,\ldots,g,\ldots,G\}$ denote the group of the $k$th column of $\bZ$. 
\item Build the diagonal matrix $\bM\bP$ of the new weights of the column of $\bZ$.
\end{enumerate}

\paragraph{Step 2: re-weighted global \code{PCAmix} step.}
\begin{enumerate}
\item The GSVD of  $\bZ$  with metrics $\bN$ on  $\R^{n}$ and $\bM\bP$ on  $\R^{p_1+m}$ gives:
$$\bZ=\bU_{\MFA}\bLambda_{\MFA}\bV_{\MFA}^\top,$$
as defined in (\ref{gsvd}). Let $r$ denote the rank of $\bZ$. 
\item The matrix of dimension $n \times r$ of the factor coordinates of the $n$ observations is:
\begin{equation}
\bF_{\MFA}=\bU_{\MFA} \bLambda_{\MFA}.
\end{equation}
\item  The matrix of dimension $(p_1+m) \times r$ of the factor coordinates of the $p_1$ quantitative variables and the $m$ levels is: 
\begin{equation}
\bA_{\MFA}^*= \bM \bV_{\MFA} \bLambda_{\MFA}. \label{eqmfa}
\end{equation}
The first $p_1$  rows contain the factor coordinates of the numerical variables and the following $m$ rows contain the factor coordinates of the levels. 
\end{enumerate}

\paragraph{Step 3: squared loading processing.}
The squared loadings are the contributions of the $p$ variables to the variance of the $r$ principal components  (columns of $\bF_{\MFA}$).   It comes from Section \ref{sec:pcamet} that the variance of the $i$th principal component $\bff_{i,{\MFA}}$ is $\mbox{Var}(\bff_{i,{\MFA}})=\| \ba_{i,{\MFA}} \|^2_{\bM\bP}$ where $\ba_{i,{\MFA}}$ is the $i$th loadings vector (column of $\bA_{\MFA}=\bV_{\MFA}\bLambda_{\MFA}$). The contribution $c_{ji,{\MFA}}$ of the variable  $\bx_j$  to the variance of the principal component $\bff_{i,{\MFA}}$ is then:
\begin{equation}
\displaystyle
\left\{
\begin{array}{ll}
\displaystyle
c_{ji,{\MFA}}=\frac{1}{\lambda_1^{(t_j)}}a_{ji,{\MFA}}^2=\frac{1}{\lambda_1^{(t_j)}}a_{ji,{\MFA}}^{*2} & \mbox{ if the variable } \bx_j  \mbox{  is numerical},\\
\displaystyle
c_{ji,{\MFA}}=\sum_{s \in I_j}\frac{1}{\lambda_1^{(t_s)}} \frac{n}{n_s} a_{si,{\MFA}}^2 =\sum_{s \in I_j}\frac{1}{\lambda_1^{(t_s)}} \frac{n_s}{n} a_{si,{\MFA}}^{*2} & \mbox{ if the variable  } \bx_j  \mbox{ is categorical},
\end{array}
\right. \label{eq:contrib3}
\end{equation}
where $ I_j$ is the set of indices of the levels of the categorical variable $\bx_j$.  Note that the contributions are no longer squared correlation or correlation ratios as previously in \code{PCArot} and \code{PCAmix}.

\begin{remark}
In general $q \leq r$  dimensions are required by the user in \code{MFAmix}. 
\end{remark}

\subsection{Graphical outputs of \code{MFAmix}} \label{sec:mfamixgraph}

The graphical outputs of \code{MFAmix} are obtained with the function \code{plot.MFAmix}. 
The standard plots (observations, numerical variables and  levels according to their factor coordinates) are interpreted with the same rules as in \code{PCAmix} (see Section~\ref{plotpcamix}) which remain true in \code{MFAmix}. The interpretation of the plot of the variables according to their squared loadings is however slightly different. Indeed, in \code{MFAmix}, squared loadings need to be interpreted as contributions and no longer as squared correlations or correlation ratios. The group structure of the variables allows to build in \code{MFAmix} new graphical outputs: plot of the groups, plot of the partial observations and plot of the partial axes.

\paragraph{Contribution of a group.}  The contribution of a variable is defined in (\ref{eq:contrib3}). The contribution of a group $g$ is therefore the sum of the contributions of all the variables of the group. The groups can then be plotted as points on a map using their contribution to the variance of the principal components.

\paragraph{Partial observations.}  The principal component map of the observations reveals the structure common  to the groups, but it is not possible to  see how each group relates to the principal component space. The visualization of an observation according to a specific group (called a partial observation)  can be achieved by projecting the dataset of each group onto this space. This is done as follows:
\begin{enumerate}
\item  For $g=1,\ldots,G$,   construct the  matrix  $\bZ^{\g}_{\pa}$ by equating to zero in $\bZ$ the values of the columns $k$ such that $t_k \neq g$. The rows of $\bZ^{\g}_\pa$ are the partial observations for the group $g$.
\item  For $g=1,\ldots,G$,  the factor coordinates of the partial observations are computed as:
\begin{equation} 
\bF^{\g}_\pa =G \; \bZ^{\g}_\pa \bM\bP \bV .
\end{equation}
This matrix contains  the coordinates of the orthogonal projections (with respect to the adjusted metric matrix $\bM\bP$) of the $n$ rows of $\bZ^{\g}_\pa$ onto the axes spanned by the columns of $\bV$ (with the number of groups $G$ as multiplying factor). This  multiplying factor comes to get the factor coordinates of an observation at the barycenter of the coordinates of its $G$ partial observations. 

\end{enumerate}
The partial observations can then be plotted as supplementary points on the principal component map of the observations. To facilitate interpretation, lines linking an observation with its $G$ partial observations are drawn on the map.

\paragraph{Partial axes.}  The \code{PCAmix} procedure is applied first to the $G$  separated data tables $\bX^{\g}$. 
The principal components $\bff_i^{\g}, i=1\ldots q$ of these separate analyses are called the partial axes. Let $\bff_{i,\MFA}$ denote the $i$th principal component of the global analysis. The link between the separated analysis and the global analysis is explored by computing correlations between the principal components of each separated study and the principal components of the global study. The correlations $r(\bff_i^{\g},\bff_{i,\MFA})$ are used as coordinates to plot the partial axes on a map.

\subsection{Prediction of PC scores with \code{predict.MFAmix}}

The $q \leq r$ principal components (PCs) are new numerical variables defined as a linear combination of the vectors $\bz_1,\ldots,\bz_{p_1+m}$ (columns  of $\bZ$). For $i=1,\ldots,q$:

$$\bff_{i,\MFA}=\bZ\bM\bP\bv_{i,{\MFA}}=\sum_{\ell=1}^{p_1}\frac{1}{\lambda_1^{(t_\ell)}}v_{\ell i,{\MFA}}\bz_j + \sum_{\ell=p_1+1}^{p_1+m}\frac{1}{\lambda_1^{(t_\ell)}}\frac{n}{n_\ell}v_{\ell i,{\MFA}}\bz_\ell. $$

It is then easy to write $\bff_{i,\MFA}$ as a linear combination of the vectors $\bx_1,\ldots,\bx_{p_1+m}$  (columns of $\bX=(\bX_1|\bG)$) where $\bG$ is the indicator matrix of the $m$ levels:
\begin{equation}
\bff_{i,\MFA}= \beta_{0i,\MFA}+\sum_{\ell=1}^{p_1+m}\beta_{\ell i,\MFA} \bx_\ell,
\label{coefmfamix}
\end{equation}
with the coefficients

\begin{align*}
 \beta_{0i,\MFA} &= -\sum_{\ell=1}^{p_1} \frac{1}{\lambda_1^{(t_\ell)}} v_{\ell i,\MFA}\frac{{\bar \bx}_\ell}{\hat{\sigma}_\ell} -\sum_{\ell=p_1+1}^{p_1+m}\frac{1}{\lambda_1^{(t_\ell)}} \frac{n}{n_\ell} v_{\ell i,\MFA} {\bar \bx},\\
\beta_{\ell i,\MFA} &= \frac{1}{\lambda_1^{(t_\ell)}} v_{\ell i,\MFA}\frac{1}{\hat{\sigma}_\ell}, \mbox{ for }  \ell=1,\ldots, p_1,\\
\beta_{\ell i,\MFA}&= \frac{1}{\lambda_1^{(t_\ell)}} \frac{n}{n_\ell} v_{\ell i,\MFA}, \mbox{ for } \ell=p_1+1,\ldots, p_1+m,
\end{align*}
where ${\bar \bx}_\ell$ and $\hat{\sigma}_\ell$ are respectively the empirical mean and the standard deviation of the column $\bx_\ell$.

The principal components are thereby written in (\ref{coefmfamix}) as a linear combination of the  original numerical variables and of the original indicator vectors of the levels of the categorical variables.
The function \code{predict.MFAmix} uses these coefficients  to predict the scores (coordinates) of new observations on the first  $q \leq r$ principal component of \code{MFAmix} (where $q$ is chosen by the user). 

\bigskip
\bigskip

\subsection{Illustration of \code{MFAmix}} \label{ex_mfamix}

Let us now illustrate the procedure \code{MFAmix} with the 4 mixed data tables available in the dataset \code{gironde}.  As introduced previously, this dataset describes $542$ municipalities on 27 variables separated into 4 groups (Employment, Housing, Services, Environment). The dataset   \code{gironde} is then a list of 4 data tables (one data table by group).

\begin{verbatim}
R> library("PCAmixdata")
R> data("gironde")
R> names(gironde)
[1] "employment"  "housing"     "services"    "environment"
\end{verbatim}

The four groups contain respectively 9, 5, 9 and 4 variables and the description of the variables of each data table is available in  Appendix~\ref{datades}.  

The function \code{MFAmix} uses three main input arguments: 
\begin{itemize}
\item[-] \code{data}: the global data frame obtained by concatenation of the separated data tables,
\item[-] \code{group}: a vector of integer with the  index of the group of each variable,
\item[-] \code{name.group}: a vector of character with the name of each group. 
\end{itemize}

\begin{verbatim}
R> dat <- cbind(gironde$employment, gironde$housing, gironde$services,
			gironde$environment) 
R> index <- c(rep(1,9), rep(2,5), rep(3,9), rep(4,4)) 
R> names <- c("employment", "housing", "services", "environment") 
R> res.mfamix <- MFAmix(data = dat, groups = index, name.groups = names,
			    ndim = 3, rename.level = TRUE, graph = FALSE)
\end{verbatim}

\begin{figure}[htb]
\begin{center}
\includegraphics[width=1\textwidth]{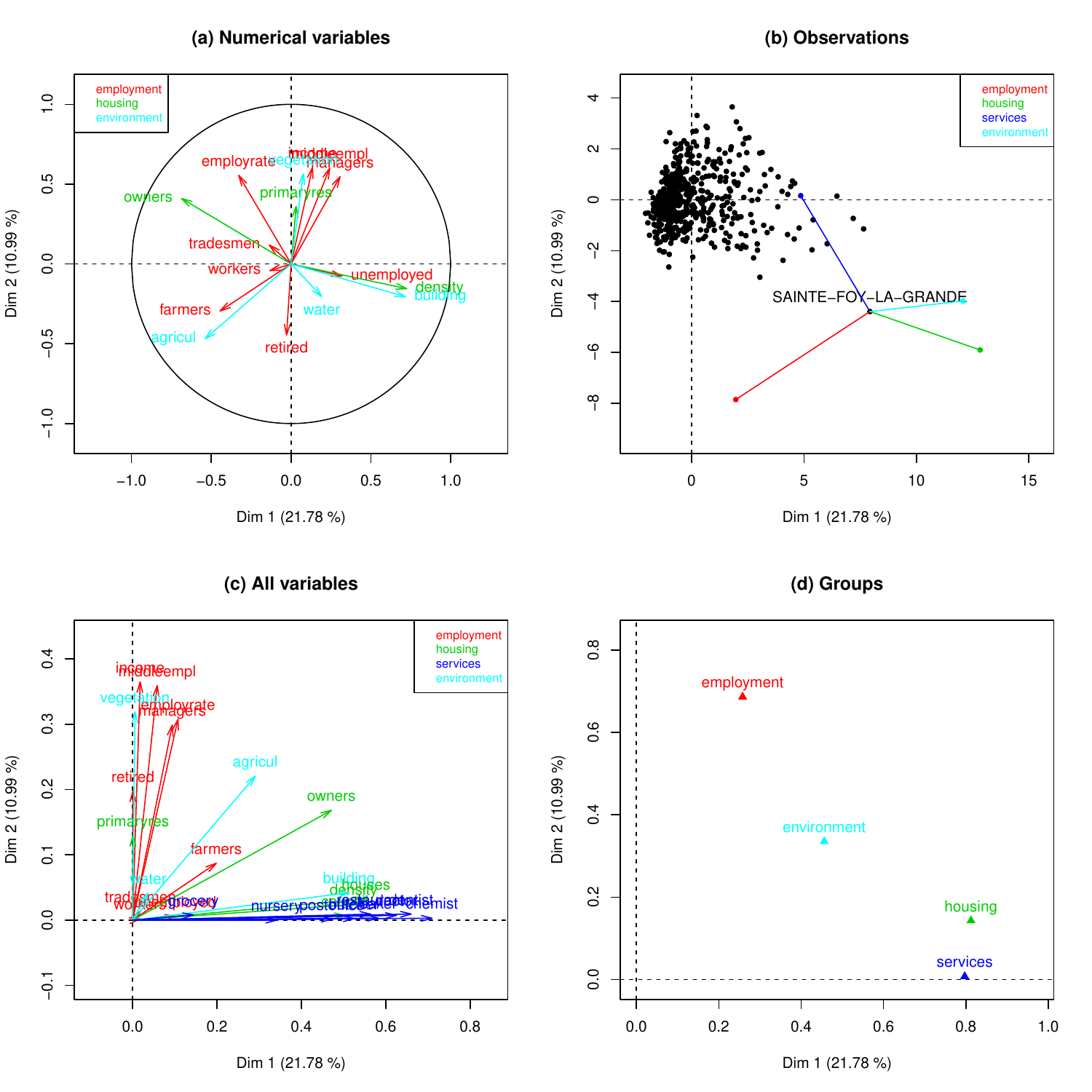} 
\caption{Some graphical outputs of MFAmix applied to the four data table of the dataset gironde.}
\label{mfamix_plot}
\end{center}
\end{figure}

The function \code{MFAmix} builds an object (of class \code{MFAmix}) which is a list with many numerical results described shortly with the \code{print} function. Here,  the number of dimensions kept in the results  is equal to 3.
The group structure of the variables gives specific graphical outputs like the four maps of Figure~\ref{mfamix_plot}.

\begin{verbatim}
R> plot(res.mfamix, choice = "cor", coloring.var = "groups", leg = TRUE,
     main = "(a) Numerical variables")
R> plot(res.mfamix, choice = "ind", partial = c("SAINTE-FOY-LA-GRANDE"), label = TRUE,
     posleg = "topright", main = "(b) Observations")
R> plot(res.mfamix, choice = "sqload", coloring.var = "groups",
     posleg = "topright", main="(c) All variables")
R> plot(res.mfamix, choice = "groups", coloring.var = "groups", main = "(d) Groups")
\end{verbatim}

Figure~\ref{mfamix_plot}(a) is the correlation circle of the 16 numerical variables, colored according to their group membership. The coordinates of the variables on this map are correlations with the principal components of \code{MFAmix}. Because this map can be difficult to read due to multiple overlaying of the names of some variables, it can be useful to look at the numerical values of the coordinates available in the object \code{res.MFAmix}.

\begin{verbatim}
R> coord.var <- res.mfamix$quanti$coord[ , 1:2]
\end{verbatim}

Table \ref{coord_var} highlights four numerical variables that are correlated (in absolute value) with the first principal component: \code{density}, \code{buildings}, \code{owners} and \code{agricul}. The municipalities  on the right hand side of the principal component map in Figure \ref{mfamix_plot}(b) have then higher values for variables \code{density} and \code{buildings}, whereas municipalities to the left  have higher values of the variables \code{owners} and \code{agric}. 

\begin{table}[htbp]
  \centering
\caption{Factor coordinates of  the variables obtained with \code{MFAmix}}	
{\small
  \label{coord_var}
\begin{tabular}{l|r|r}
\hline
  & dim 1 & dim 2\\
\hline
farmers & -0.45 & -0.30\\
\hline
tradesmen & -0.14 & 0.12\\
\hline
\textbf{managers} & 0.31 &\textbf{ 0.55}\\
\hline
workers & -0.13 & -0.04\\
\hline
unemployed & 0.32 & -0.08\\
\hline
\textbf{middleempl} & 0.24 & \textbf{0.60}\\
\hline
retired & -0.03 & -0.44\\
\hline
employrate & -0.33 & 0.55\\
\hline
\textbf{income} & 0.13 & \textbf{0.60}\\
\hline
\textbf{density} & \textbf{0.72} & -0.15\\
\hline
primaryres & 0.03 & 0.36\\
\hline
\textbf{owners} & \textbf{-0.69} & 0.41\\
\hline
\textbf{building} & \textbf{0.72} & -0.21\\
\hline
water & 0.19 & -0.20\\
\hline
\textbf{vegetation} & 0.08 & 0.56\\
\hline
\textbf{agricul }& \textbf{-0.54} & -0.47\\
\hline
\end{tabular}
}
\end{table}

To interpret the position of the municipalities at the top and bottom of Figure \ref{mfamix_plot}(b), the coordinates of the variables in the second dimension are useful. Table \ref{coord_var} highlights four numerical variables that are correlated with the second principal component:  \code{managers}, \code{middleempl}, \code{employrate}, \code{income} and \code{vegetation}. The position (top or bottom) of the municipalities on the principal component map can then be interpreted with these variables.

For example, Figure~\ref{mfamix_plot}(b)  shows the municipality of  \code{SAINTE-FOY-LA-GRANDE}  plotted with its 4 partial representations (the four colored points linked to it with a line). The position of this municipality on the right of the map suggests a municipality with higher density of population, higher proportion of buildings, less owners and less agricultural land. Its position at the bottom of the map suggests smaller values on 4 variables of the group \code{employment} (\code{managers}, \code{middleempl},\code{employrate},\code{income}) and smaller values on the variable \code{vegetation} of the group \code{environment}.  

\begin{figure}[htb]
\begin{center}
\includegraphics[width=0.7\textwidth]{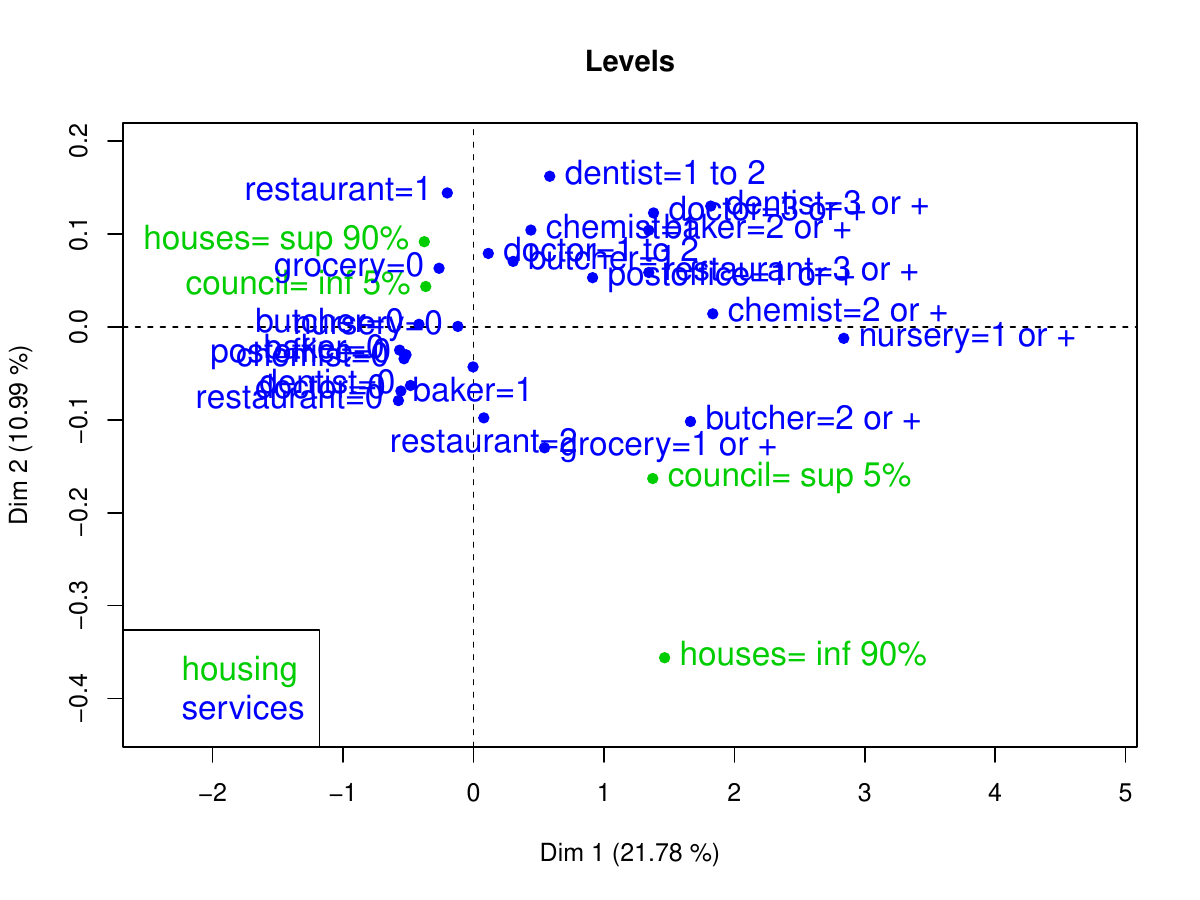} 
\caption{Plot of the levels of the 10 categorical variables after applying MFAmix.}
\label{mfamix_plot2}
\end{center}
\end{figure}

Now we come to the 9 categorical variables of the group  \code{services}. These variables naturally do not appear in the correlation circle, but do appear in Figure ~\ref{mfamix_plot}(c) where all the variables are plotted according to their contributions to the principal components. This map shows that all the variables of the group \code{services} (\code{dentist}, \code{dentist}, \code{nursery},...) contribute strongly to the first principal component. However it is not possible to know in which way. For instance does the municipality  \code{SAINTE-FOY-LA-GRANDE} which has a high score on the first principal component have more or less services than others? This information is given in Figure~\ref{mfamix_plot2} where the levels of the categorical variables are plotted.

\begin{verbatim}
R> plot(res.mfamix, choice = "levels", coloring.var = "groups",
     posleg = "bottomleft", main = "Levels", cex = 1.3, cex.leg = 1.3, xlim = c(-2,4))
\end{verbatim}

The level map can be used with the barycentric property to interpret the map of the municipalities given Figure \ref{mfamix_plot}(b): the municipalities on the right are provided with more services than those on the left. The municipalities to the bottom right of the map (like  \code{SAINTE-FOY-LA-GRANDE}) have more likely a smaller proportion of houses. 

In summary, the municipality \code{SAINTE-FOY-LA-GRANDE} is a municipality with a good level of services, but with a fairly stagnant employment market and whose inhabitants are more likely to live in apartments than in other municipalities.  

The last map Figure \ref{mfamix_plot}(d) is the plot of the groups according to their contributions to the first two principal components. This map confirms the previous interpretations of the principal components of \code{MFAmix} and the impact of the groups \code{services} and \code{housing}  on the first dimension as well as the impact of the group \code{employment} on the second dimension. 

\begin{figure}[htb]
\begin{center}
\includegraphics[width=0.7\textwidth]{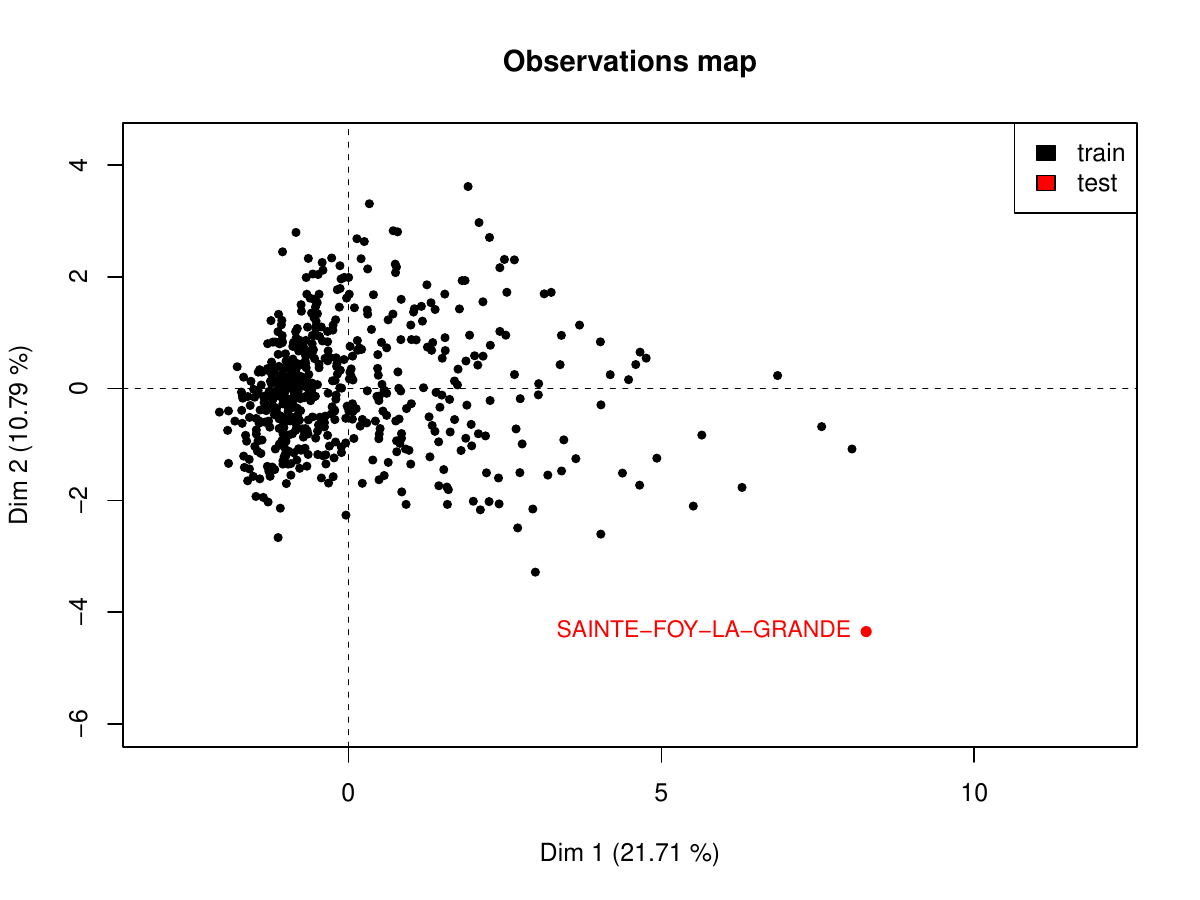} 
\caption{The municipality   SAINTE-FOY-LA-GRANDE is plotted in supplementary in the graphical output of MFAmix.}
\label{mfamix_plot3}
\end{center}
\end{figure}

\paragraph{Predicted scores for new observations.} 
The scores of new observations can be obtained with the \code{predict.MFAmix} function. The municipality \code{SAINTE-FOY-LA-GRANDE} for instance can be considered as supplementary and plotted as an illustrative observation (test sample) on the map given in Figure~\ref{mfamix_plot3} obtained with the $n-1$ remaining municipalities (training sample).
\begin{verbatim}
R> sel <- which(rownames(dat) == "SAINTE-FOY-LA-GRANDE")
R> res.mfamix <- MFAmix(data = dat[-sel,], groups = index,
                   name.groups = names, rename.level = TRUE, graph = FALSE)
R> pred <- predict(res.mfamix, dat[sel, , drop=FALSE])
\end{verbatim}

\paragraph{Supplementary groups.}  The \code{supvar.MFAmix} function calculates the coordinates of supplementary groups of variables on the maps of \code{MFAmix}. Let us for instance apply \code{MFAmix} with three groups (\code{employment},  \code{services}, \code{environment}) and add the group \code{housing} as a supplement.
\begin{verbatim}
R> dat <- cbind(gironde$employment, gironde$services, gironde$environment) 
R> names <- c("employment", "services", "environment") 
R> mfa <-MFAmix(data = dat, groups = c(rep(1,9), rep(2,9), rep(3,4)), 
                   name.groups = names, rename.level =T RUE, graph = FALSE)
R> mfa.sup <- supvar(mfa, data.sup = gironde$housing, groups.sup = rep(1,5),
                name.groups.sup = "housing.sup", rename.level = TRUE)
\end{verbatim}

\begin{figure}[htb]
\begin{center}
\includegraphics[width=1\textwidth]{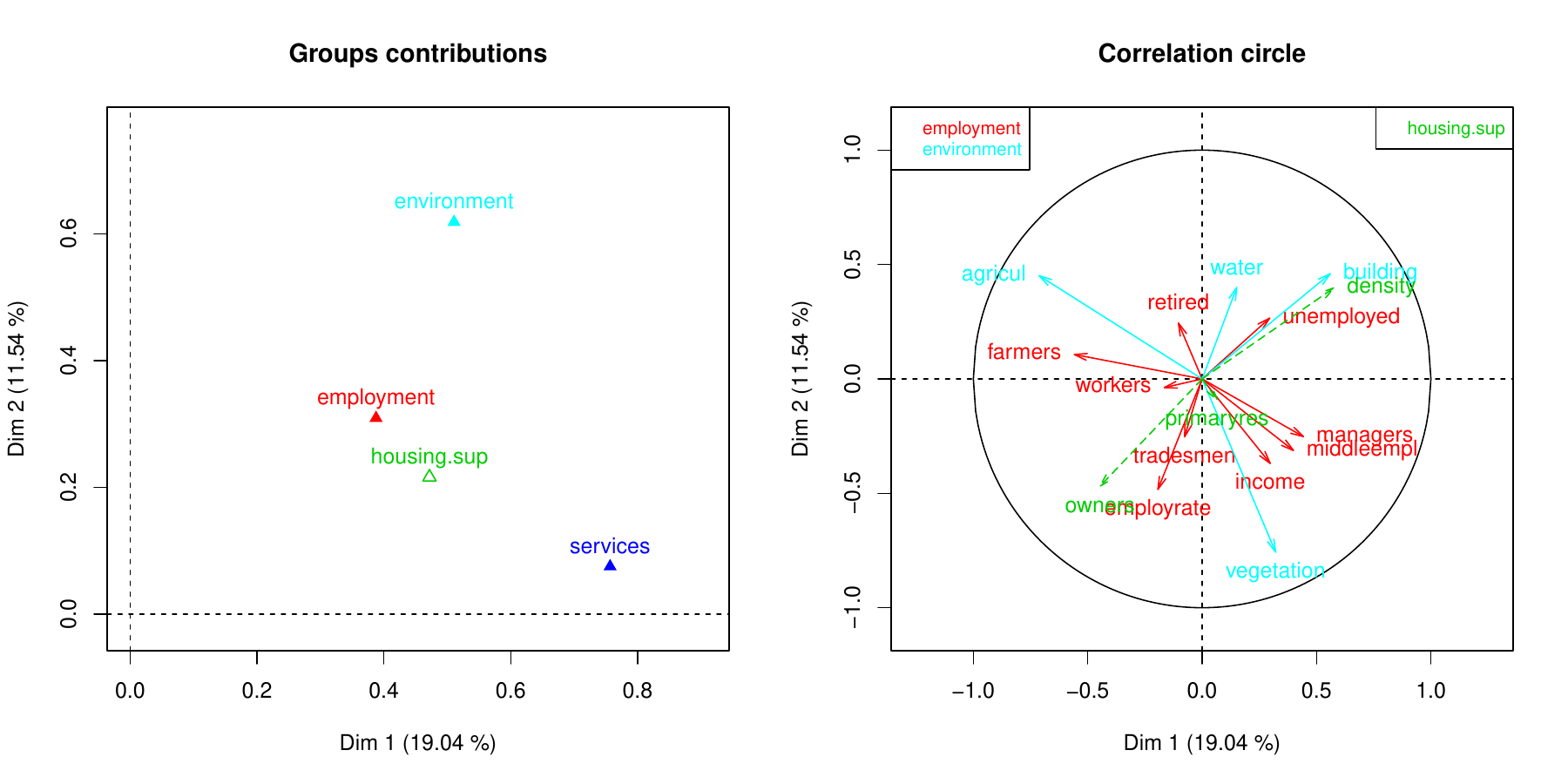} 
\caption{The group  housing is plotted as supplementary in the graphical outputs of MFAmix.}
\label{mfamix_plot4}
\end{center}
\end{figure}

The group \code{housing} is then plotted as supplementary on the maps of \code{MFAmix}, see Figure \ref{mfamix_plot4}.

\begin{verbatim}
R> plot(mfa.sup, choice = "groups", coloring.var = "groups",
     		col.groups = c(2,4,5), col.groups.sup = 3)
R> plot(mfa.sup,choice="cor", coloring.var = "groups",
     		col.groups = c(2,4,5), col.groups.sup = 3)
\end{verbatim}

\section{Concluding remarks}\label{conclu}

In this paper, the multivariate analysis methods implemented in the \proglang{R}  package \code{PCAmixdata} are presented in such a way that the theoretical details can be read separately from the  \proglang{R} examples.  Therefore,  users interested in the practical aspects of the methods \code{PCAmix}, \code{PCArot} and \code{MFAmix} can reproduce the \proglang{R} code provided after each theoretical section, either with the dataset  \code{gironde} (available in the package) or with their own data. Keys are also provided for the interpretation of most numerical results and graphical outputs. 

The definition of multivariate analysis methods to mixed data is important in practice and is sometimes  neglected in statistical literature and software. Research and implementation work remain to be done in this sense. For instance, the development of a method of linear discriminant analysis  compatible with mixed data is currently under investigation. Moreover, extension of orthogonal rotation to the principal component of \code{MFAmix} could be done in the same spirit as \code{PCArot}, because \code{MFAmix} is a re-weighted  general \code{PCAmix} analysis, this implementation should not require too many theoretical developments. 

The package  \code{PCAmixdata} handles missing data with a very simple approach where missing values are replaced by mean values for numerical variables and by zeros in the indicator matrix for the categorical variables. Of course more relevant methods like the method proposed by \cite{audigier2016principal} and implemented in the \proglang{R} package \pkg{missMDA} could be used to complete the missing values.

\section*{Acknowledgments}
The authors warmly thank the anonymous reviewers who contribute to greatly improve the manuscript.

\appendix
\appendixpage
\section[]{The dataset \code{gironde}}\label{datades}

Table \ref{recap_var} provides the description of all the numerical and categorical variables of the  \code{gironde} dataset.
\begin{table}[htbp]
\caption{Description of variables of the  \code{gironde} dataset}	
{\small
  \centering
  \label{recap_var}
    \begin{tabular}{|l|l|l|l|}
    \hline
    \textbf{R\_Names} & \textbf{Description} & \textbf{Group} & \textbf{Data type} \\
    \hline
    farmers & Percentage of farmers & employment & Num \\
    tradesmen & Percentage of tradesmen and shopkeepers & employment & Num \\
    managers & Percentage of managers and executives & employment & Num \\
    workers & Percentage of workers and employees & employment & Num \\
    unemployed & Percentage of unemployed workers & employment & Num \\
    middleemp & Percentage of middle-range employees & employment & Num \\
    retired & Percentage of retired people & employment & Num \\
    employrate & employment rate & employment & Num \\
    income & Average income & employment & Num \\
    \hline
    density & Population density & housing & Num \\
    primaryres & Percentage of primary residences & housing & Num \\
    houses & Percentage of houses & housing & Categ \\
    owners & Percentage of home owners living in their primary residence & housing & Num \\
    council & Percentage of council housing & housing & Categ \\
    \hline
    butcher & Number of butchers & services & Categ \\
    baker & Number of bakers & services & Categ \\
    postoffice & Number of post offices & services & Categ \\
    dentist & Number of dentists & services & Categ \\
    grocery & Number of grocery stores & services & Categ \\
    nursery & Number of child care day nurseries & services & Categ \\
    doctor & Number of doctors & services & Categ \\
    chemist & Number of chemists & services & Categ \\
    restaurant & Number of restaurants & services & Categ \\
     \hline
    building & Percentage of buildings & environment & Num \\
    water & Percentage of water & environment & Num \\
    vegetation & Percentage of vegetation & environment & Num \\
    agricul & Percentage of agricultural land & environment & Num \\
    \hline
    \end{tabular}
  }
\end{table}


\section[]{The iterative optimization step of \code{PCArot}}\label{Appendix:pcarot}

Let $\tbU_q$  (resp. $\tbA_q$) denote the matrix of the first  $q$ columns of $\tbU$ (resp. $\tbA= {\tilde \bLambda}  {\tilde \bV}$).
\begin{enumerate}
\item Initialization: $\tbU_{\mbox{\tiny rot}}=\tbU_q$ and $\tbA_{\mbox{\tiny rot}}=\tbA_q$.
\item For each pair of  dimensions $(l,t)$, i.e., for $l=1,\ldots,q-1$ and $t=(l+1),\ldots,q$: 
\begin{itemize}
\item[$\hookrightarrow$]  calculate the angle of rotation $ \theta={\psi}/{4} $ with:
\begin{equation}
\displaystyle
\psi=   \left \{
   \begin{array}{ rll}
   \displaystyle   \text{arcos}\left(\frac{h}{\sqrt{g^2+h^2}} \right) & \text{if} & g \geq 0,\\
   \displaystyle   -\text{arcos}\left(\frac{b}{\sqrt{g^2+h^2}}\right)  & \text{if} & g \leq 0,
   \end{array}
   \right .
\label{psi}
\end{equation}
where $g$ and $h$ are  given by:
\begin{align}
 g&= 2p \sum_{j=1}^p {\alpha_j} {\beta_j} -2\sum_{j=1}^p {\alpha_j} \sum_{j=1}^p {\beta_j},\\
 h&= p \sum_{j=1}^p ({\alpha_j}^2-{\beta_j}^2) -\left( \sum_{j=1}^p {\alpha_j}\right)^2 + \left( \sum_{j=1}^p {\beta_j}\right)^2,
\end{align}
with $p$ the total number of variables, and $\alpha_j$ and $\beta_j$  defined by:
\begin{equation} 
\alpha_j= \sum_{s \in I_j} (\ta_{sl,\mbox{\tiny rot}}^2-\ta_{st,\mbox{\tiny rot}}^2) ~~~\mbox{and}~~~ {\beta_j}=2\sum_{s \in I_j} \ta_{sl,\mbox{\tiny rot}}\ta_{st,\mbox{\tiny rot}}\ .
\end{equation}
Here, $I_j$ is the set of row indices of  $\tbA_{\mbox{\tiny rot}}$ associated with the levels of the variable $j$ in the categorical case and $I_j=\{j\}$ in the numerical case. 
\item[$\hookrightarrow$] calculate the corresponding matrix of planar rotation $
\bT_2=
\left[
\begin{array}{cc}
\text{cos } \theta & -\text{sin } \theta \\
\text{sin } \theta & \text{cos } \theta
\end{array}
\right]
$,
\item[$\hookrightarrow$] update the matrices  $\tbU_{\mbox{\tiny rot}}$ and  $\tbA_{\mbox{\tiny rot}}$  by rotation of their $l$-th and $t$-th columns. 
\end{itemize}
\item Repeat the previous step until the $q(q-1)/2$ successive rotations provide an angle of rotation $ \theta$ equal to zero. 
\end{enumerate}

\section[]{Equivalence between (\ref{varipcamix}) and (\ref{varipcamix2})}
\label{appendix:proof1}
We know from (\ref{astar}) that:
$$\bA^* =\bM \bA=\bM \bV \bLambda.$$
Moreover we know from (\ref{back}) that $\bV=\bM^{-1/2}{\tilde \bV}$ and that $\bLambda={\tilde \bLambda}$. It gives then:
\begin{equation}
\bA^* =\bM^{1/2}{\tilde \bV} {\tilde \bLambda}=\bM^{1/2}{\tilde \bA}.
\label{adac}
\end{equation}
Note that in \cite{chavent2012orthogonal} we wrote $\bA^*=\bM^{-1/2}{\tilde \bA}$ which was a mistake. It gives also 
\begin{equation}
\bA= \bM^{-1/2}{\tilde \bV}{\tilde \bLambda}=\bM^{-1/2}{\tilde \bA}.
\label{adac2}
\end{equation}

We deduce easily from (\ref{adac2}) that:

\begin{equation}
\left\{
\begin{array}{ll}
c_{ji}=a_{ji}^2 ={\tilde a}_{ji}^2= {\tilde c}_{ji} & \mbox{ if the variable }\bx_ j   \mbox{  is numerical},\\
c_{ji}=\sum_{s \in I_j} \frac{n}{n_s} a_{si}^2=\sum_{s \in I_j}  {\tilde a}_{si}^2= {\tilde c}_{ji}  & \mbox{ if the variable  } \bx_ j  \mbox{ is categorical},
\end{array}
\right.
\end{equation}

\section[]{Proof of (\ref{defFrot})}\label{appendix:proof}

The $q \times q$ rotation matrix $\bT$ is such that 
\begin{equation}
 \tbU_{\mbox{\tiny rot}}=\tbU_q \bT.
 \end{equation}
By definition of  $\tbU_q$, we have $\tbU_q^\top\tbU_q=\mathbb{I}_q$. It gives (\ref{defT}).  
By definition, $\tbF_{\mbox{\tiny rot}}=\tbU_{\mbox{\tiny rot}}\bLambda_{\mbox{\tiny rot}}$. It gives  $\tbF_{\mbox{\tiny rot}}=\tbU_q\bT\bLambda_{\mbox{\tiny rot}}$. The SVD decomposition $\tbZ=\tbU \tbLambda \tbV^\top$ gives $\tbU_q=\tbZ \tbV_q\tbLambda_q^{-1}$. Then $\tbF_{\mbox{\tiny rot}}=\tbZ \tbV_q\tbLambda_q^{-1}\bT\bLambda_{\mbox{\tiny rot}}$. With $\tbF_{\mbox{\tiny rot}}=\bN^{1/2}\bF_{\mbox{\tiny rot}}$ and $\tbZ=\bN^{1/2}\bZ \bM^{1/2}$, it gives (\ref{defFrot}) and (\ref{defVrot}).

%
\bibliographystyle{apalike}
\bibliography{pcamixdata}

\begin{thebibliography}{}

\bibitem[Abdi, 2007]{abdi2007singular}
Abdi, H. (2007).
\newblock Singular value decomposition (svd) and generalized singular value
  decomposition.
\newblock {\em Encyclopedia of measurement and statistics}, pages 907--912.

\bibitem[Abdi et~al., 2013]{abdi2013multiple}
Abdi, H., Williams, L.~J., and Valentin, D. (2013).
\newblock Multiple factor analysis: Principal component analysis for multitable
  and multiblock data sets.
\newblock {\em Wiley Interdisciplinary Reviews: Computational Statistics},
  5(2):149--179.

\bibitem[Audigier et~al., 2016]{audigier2016principal}
Audigier, V., Husson, F., and Josse, J. (2016).
\newblock A principal component method to impute missing values for mixed data.
\newblock {\em Advances in Data Analysis and Classification}, 10(1):5--26.

\bibitem[Beaton et~al., 2014]{pExPosition}
Beaton, D., Fatt, C. R.~C., and Abdi, H. (2014).
\newblock An {ExPosition} of multivariate analysis with the singular value
  decomposition in {R}.
\newblock {\em Computational Statistics and Data Analysis}, 72(0):176 -- 189.

\bibitem[B{\'e}cue-Bertaut and Pag{\`e}s, 2008]{becue2008}
B{\'e}cue-Bertaut, M. and Pag{\`e}s, J. (2008).
\newblock Multiple factor analysis and clustering of a mixture of quantitative,
  categorical and frequency data.
\newblock {\em Computational Statistics and Data Analysis}, 52(6):3255--3268.

\bibitem[Chavent et~al., 2017]{pcamixdatar}
Chavent, M., Kuentz-Simonet, V., Labenne, A., Liquet, B., and Saracco, J.
  (2017).
\newblock {\em \pkg{PCAmixdata}: Multivariate Analysis of Mixed Data}.
\newblock \proglang{R}~package version~3-1.

\bibitem[Chavent et~al., 2012]{chavent2012orthogonal}
Chavent, M., Kuentz-Simonet, V., and Saracco, J. (2012).
\newblock Orthogonal rotation in pcamix.
\newblock {\em Advances in Data Analysis and Classification}, 6(2):131--146.

\bibitem[{de Leeuw} and Mair, 2009]{homals}
{de Leeuw}, J. and Mair, P. (2009).
\newblock Gifi methods for optimal scaling in {R}: The package {homals}.
\newblock {\em Journal of Statistical Software}, 31(4):1--20.

\bibitem[de~Leeuw and van Rijckevorsel, 1980]{deleeuw80}
de~Leeuw, J. and van Rijckevorsel, J. L.~A. (1980).
\newblock Homals and princals, some generalizations of principal components
  analysis.
\newblock In Diday, E. and al., editors, {\em Data analysis and informatics
  II}, pages 231--242. Elsevier Science Publishers, Amsterdam.

\bibitem[Dray and Dufour, 2007]{dray2007ade4}
Dray, S. and Dufour, A.-B. (2007).
\newblock The \pkg{ade4} package: Implementing the duality diagram for
  ecologists.
\newblock {\em Journal of Statistical Software}, 22(4):1--20.

\bibitem[Dray et~al., 2017]{rade4}
Dray, S., Dufour, A.-B., Thioulouse, J., et~al. (2017).
\newblock {\em \pkg{ade4}: Analysis of Ecological Data : Exploratory and
  Euclidean Methods in Environmental Sciences}.
\newblock \proglang{R}~package version~1.7-8.

\bibitem[Escofier and Pag{\`e}s, 1994]{escofier1994multiple}
Escofier, B. and Pag{\`e}s, J. (1994).
\newblock Multiple factor analysis (\pkg{AFMULT} package).
\newblock {\em Computational Statistics \& Data Analysis}, 18(1):121--140.

\bibitem[Hill and Smith, 1976]{hill1976principal}
Hill, M. and Smith, A. (1976).
\newblock Principal component analysis of taxonomic data with multi-state
  discrete characters.
\newblock {\em Taxon}, 25(2/3):249--255.

\bibitem[Husson et~al., 2017]{rFactoMineR}
Husson, F., Josse, J., L{\^e}, S., and Mazet, J. (2017).
\newblock {\em \pkg{FactoMineR}: Multivariate Exploratory Data Analysis and
  Data Mining}.
\newblock \proglang{R}~package version~1.38.

\bibitem[Kaiser, 1958]{kaiser1958varimax}
Kaiser, H.~F. (1958).
\newblock The varimax criterion for analytic rotation in factor analysis.
\newblock {\em Psychometrika}, 23(3):187--200.

\bibitem[Kiers, 1991]{kiers1991simple}
Kiers, H.~A. (1991).
\newblock Simple structure in component analysis techniques for mixtures of
  qualitative and quantitative variables.
\newblock {\em Psychometrika}, 56(2):197--212.

\bibitem[L{\^e} et~al., 2008]{le2008factominer}
L{\^e}, S., Josse, J., and Husson, F. (2008).
\newblock \pkg{FactoMineR}: an r package for multivariate analysis.
\newblock {\em Journal of Statistical Software}, 25(1):1--18.

\bibitem[Mair et~al., 2019]{gifi}
Mair, P., Leeuw, J.~D., and Groenen, P. J.~F. (2019).
\newblock {\em \pkg{Gifi}: Multivariate Analysis with Optimal Scaling}.
\newblock \proglang{R}~package version~0.3-9.

\bibitem[Pag{\`e}s, 2004]{pages2004analyse}
Pag{\`e}s, J. (2004).
\newblock Analyse factorielle de donn{\'e}es mixtes.
\newblock {\em Revue de Statistique Appliqu{\'e}e}, 52(4):93--111.

\bibitem[{\proglang{R} Core Team}, 2017]{team2017r}
{\proglang{R} Core Team} (2017).
\newblock {\em \proglang{R}: A Language and Environment for Statistical
  Computing}.
\newblock R Foundation for Statistical Computing, Vienna, Austria.

\end{thebibliography}

%
%
%


\end{document}